\newcommand{\teff}{T$_{\mathrm{eff}}$}
\newcommand{\logg}{log $g$}

\newcommand{\ps}{$\rm{\Delta}$P}
\newcommand{\fsep}{$\rm{\Delta\nu}$}

\newcommand{\project}[1]{\textsl{#1}}
\newcommand{\gaia}{\project{Gaia}}

\newcommand{\michz}{$\rm{\mu}$Hz}

\documentclass[a4paper,fuseAMS,leqn,usenatbib]{mnras}

\usepackage{graphicx}
\usepackage{amssymb}
\usepackage{amsmath}
\usepackage{times}
\usepackage{subfigure}
\usepackage[T1]{fontenc}
\usepackage{ae,aecompl}

\title[Photometric Red Clump Stars Catalog]{From the Inner to Outer Milky Way: A Photometric Sample of 2.6 Million Red Clump Stars}
 \author[M.~Lucey et. al.]{Madeline~Lucey$^{1}$\thanks{E-mail: m\_lucey@utexas.edu}, Yuan-Sen Ting$^{2,3,4,5}$\thanks{NASA Hubble Fellow}, Nesar S. Ramachandra$^{6,7}$, Keith Hawkins$^{1}$\\
$^{1}$Department of Astronomy, the University of Texas at Austin, 2515 Speedway Boulevard, Austin, TX 78712, USA \\
$^{2}$Institute for Advanced Study, Princeton, NJ 08540, USA\\
$^{3}$Department of Astrophysical Sciences, Princeton University, Princeton, NJ 08544, USA\\
$^{4}$Observatories of the Carnegie Institution of Washington, 813 Santa Barbara Street, Pasadena, CA 91101, USA \\
$^{5}$Research School of Astronomy \& Astrophysics, Australian National University, Cotter Rd., Weston, ACT 2611, Australia \\
$^{6}$High Energy Physics Division, Argonne National Laboratory, Lemont, IL 60439, USA \\
$^{7}$Kavli Institute for Cosmological Physics, University of Chicago, 5640 South Ellis Avenue, Chicago, IL 60637, USA}

\date{Accepted .... Received ...; in original form ...}

\pubyear{2020}

\begin{document}
\label{firstpage}
\pagerange{\pageref{firstpage}--\pageref{lastpage}}
\maketitle

\begin{abstract}
Large pristine samples of red clump stars are highly sought after given that they are standard candles and give precise distances even at large distances. However, it is difficult to cleanly select red clumps stars because they can have the same \teff\ and \logg\ as red giant branch stars. Recently, it was shown that the asteroseismic parameters, \ps\ and \fsep, which are used to accurately select red clump stars, can be derived from spectra using the change in the surface carbon to nitrogen ratio ([C/N]) caused by mixing during the red giant branch. This change in [C/N] can also impact the spectral energy distribution. In this study, we predict the \ps, \fsep, \teff\ and \logg\ using 2MASS, AllWISE, \gaia, and Pan-STARRS data in order to select a clean sample of red clump stars. We achieve a contamination rate of $\sim$20\%, equivalent to what is achieved when selecting from \teff\ and \logg\ derived from low resolution spectra.  Finally, we present two red clump samples. One sample has a contamination rate of $\sim$ 20\% and $\sim$ 405,000 red clump stars. The other has a contamination of $\sim$ 33\% and $\sim$ 2.6 million red clump stars which includes $\sim$ 75,000 stars at distances $>$ 10 kpc. For |b|>30 degrees we find $\sim$ 15,000 stars with contamination rate of $\sim$ 9\%. The scientific potential of this catalog for studying the structure and formation history of the Galaxy is vast given that it includes millions of precise distances to stars in the inner bulge and distant halo where astrometric distances are imprecise.

\end{abstract}

\begin{keywords}
stars: distances, techniques: photometric, stars: evolution
\end{keywords}

\section{Introduction}
Distances are one of the most important and yet hardest measurements to make in astronomy. Distances to stars are especially important for studying the formation and the chemodynamical evolution of the Galaxy.  The \gaia\ mission aims to provide parallax measurements for billions of stars in order to produce a 3D map of the Galaxy \citep{Gaia2016}. However, the \gaia\ mission will be plagued by large uncertainties in the distant Galaxy. Through simple uncertainty propagation, the uncertainty on distance derived from parallax scales as distance squared\footnote{
$\sigma_{d}=\sqrt{\left(\frac{\partial d}{\partial \varpi}\right)^2\sigma_{\varpi}^2} = \frac{\sigma_{\varpi}}{\varpi^2} = \sigma_{\varpi}d^2$
where $d$ is the distance, $\varpi$ is the parallax and $d=1/\varpi$}. However, standard candles, such as red clump (RC) stars, where the distance is derived using the distance modulus have uncertainties that are linear with distance\footnote{
$\sigma_{d}=\sqrt{\left(\frac{\partial d}{\partial m}\right)^2\sigma_{m}^2} =\sigma_{m} \ln{\left(10\right)}10^{(m-M+5)/5} = \sigma_{m} \ln{\left(10\right)}d$
where $m$ is apparent magnitude, $M$ is the absolute magnitude and $d = 10^{(m-M+5)/5}$}. With high precision photometry, RC stars can provide distances with uncertainties < 6\% at distances up to $\sim$ 10 kpc \citep{Bovy2014,Hawkins2017,Ting2018}. Therefore, standard candles can provide more precise distances for objects in the distant Galaxy than parallaxes and are an excellent complement to the \textit{Gaia} catalog. 

In this context, RC stars are low mass stars (<2.5 $\rm{M_{\odot}}$) that have undergone the helium flash to ignite helium burning in their cores. The electron-degenerate cores prior to helium ignition causes RC stars to have roughly the same core mass. Therefore, the luminosity of RC stars is only weakly dependent on mass and metallicity \citep[][and references therein]{Girardi2016}. Beginning in the late 1990s, the nearly constant luminosity of RC stars has been used to determine distances, making RC stars a commonly used standard candle. Selecting from color-magnitude diagrams, \citet{Stanek1997} used RC stars to constrain the the shape and orientation of the Galactic bar. \citet{Stanek1998} used a similar technique to determine the distance to M31 using RC stars. However, there are many sciences cases when a more pristine sample of RC stars is desired which can be difficult to select. \citet{Bovy2014} achieved a contamination rate of $\sim$ 10\% using effective temperature (\teff) and  surface gravity (\logg) from the high-resolution spectra of the Apache Point Observatory Galactic Evolution Experiment (APOGEE) survey \citep{Majewski2017}. This contamination is from red giant branch (RGB) stars which have inert helium cores with a hydrogen burning shell but can have similar \teff\ and \logg\ as RC stars \citep{Iben1968}.

Asteroseismology has proven to be the most accurate method for selecting RC stars \citep{Montalban2010,Bedding2011,Mosser2011,Mosser2012,Stello2013,Mosser2014,Pinsonneault2014,Vrard2016,Elsworth2017}.  The average large frequency spacing (\fsep) goes as the square root of the mean stellar density \citep{Chaplin2013}. As such, it has been shown that the distribution of \fsep\ values for RGB and RC stars are distinct \citep{Miglio2009,Mosser2010}. RC stars are more constrained in this space with \fsep\ < 5 $\rm{\mu}$Hz. While, RGB stars can have \fsep\ > 20 $\rm{\mu}$Hz. Furthermore, evolved stars have been shown to have coupling between the gravity waves in the dense radiative core (g-modes) and the acoustic waves in the envelope, p-modes \citep{Beck2011}. RC stars have a lower core density than RGB stars which causes stronger coupling between g- and p-modes. Therefore, RC and RGB clearly separate in the period spacing (\ps) of these mixed modes with RC stars typically having \ps\ > 200 s and RGB stars having \ps\ <100 s \citep{Bedding2011,Mosser2011,Stello2013,Mosser2014}. However, \ps\ is a difficult measurement and has only been measured for a fraction of the \textit{Kepler} and CoRoT samples \citep{Girardi2016}.

Recently, \citet{Hawkins2018} showed the \ps\ and the \fsep\ spacing can be inferred from a stellar spectrum. It has long been thought that the carbon to nitrogen ratio, [C/N], would be different in RC stars relative to RGB stars because of mixing that occurs along the upper RGB phase \citep{Martell2008,Masseron2015,Masseron2017,Masseron2017b}.  \citet{Hawkins2018} showed the carbon and nitrogen bands in spectra from the APOGEE survey can be used to infer the \ps\ and \fsep\ and therefore, used to select RC stars. Building off of this work, \citet{Ting2018} presented a catalog of $\sim$ 100,000 RC stars from the APOGEE and LAMOST catalogs. However, > 70\% of that sample is within 3 kpc of the Sun where \textit{Gaia} gives more precise distances \citep{Ting2018}. This represents one of the largest limitations to recent spectroscopic surveys. Spectra require more flux than photometry and therefore high resolution (R = $\lambda/\Delta \lambda$ $\sim$ 22,500) and high signal-to-noise (S/N) data can only be achieved for nearby bright stars as. For example, the magnitude limit of the APOGEE survey is H= 13.8 mag \citep{Majewski2017}. Given that the RC absolute magnitude in the H-band is -1.46 mag \citep{Hawkins2017}  the APOGEE survey is limited to RC stars within 12 kpc, assuming not extiction. In the case of the LAMOST survey, which has much lower resolution (R $\sim$ 1800), the magnitude limit is r=18.5 mag which corresponds to a RC distance of $\sim$40 kpc given the RC absolute magnitude in the r-band of 0.55 mag \citep{Ruiz-Dern2018}. However, these observation require long exposure times (5400 s) and therefore are number limited. Photometric surveys are able to achieve fainter magnitudes for much greater number of stars. For example, the RC catalog derived using LAMOST spectra has $\sim$ 2,000 RC stars with distances >10 kpc \citep{Ting2018} while we find $\sim$ 75,000 RC stars with distances >10 kpc using photometric surveys.

In this work we aim to make use of the vast amount of available photometry to obtain the largest and most distant sample of RC stars yet. In Section \ref{sec:dataall} we describe the photometric selection we use to create the spectral energy distribution (SED). In Section \ref{sec:method}, we describe the method we develop to infer the \teff, \logg, \ps\ and \fsep\ from the SEDs using a neural network based method and how we use those parameters to select a sample of RC stars.  In Section \ref{sec:sample}, we present the catalog of RC stars. In Section \ref{sec:interpret}, we demonstrate the interpretability of the neural network and try to confirm the physical mechanism used for the inference. Finally, we summarize the results in Section \ref{sec:summary}.

\section{Data} \label{sec:dataall}

\begin{figure*}
    \centering
    \includegraphics[width=\linewidth]{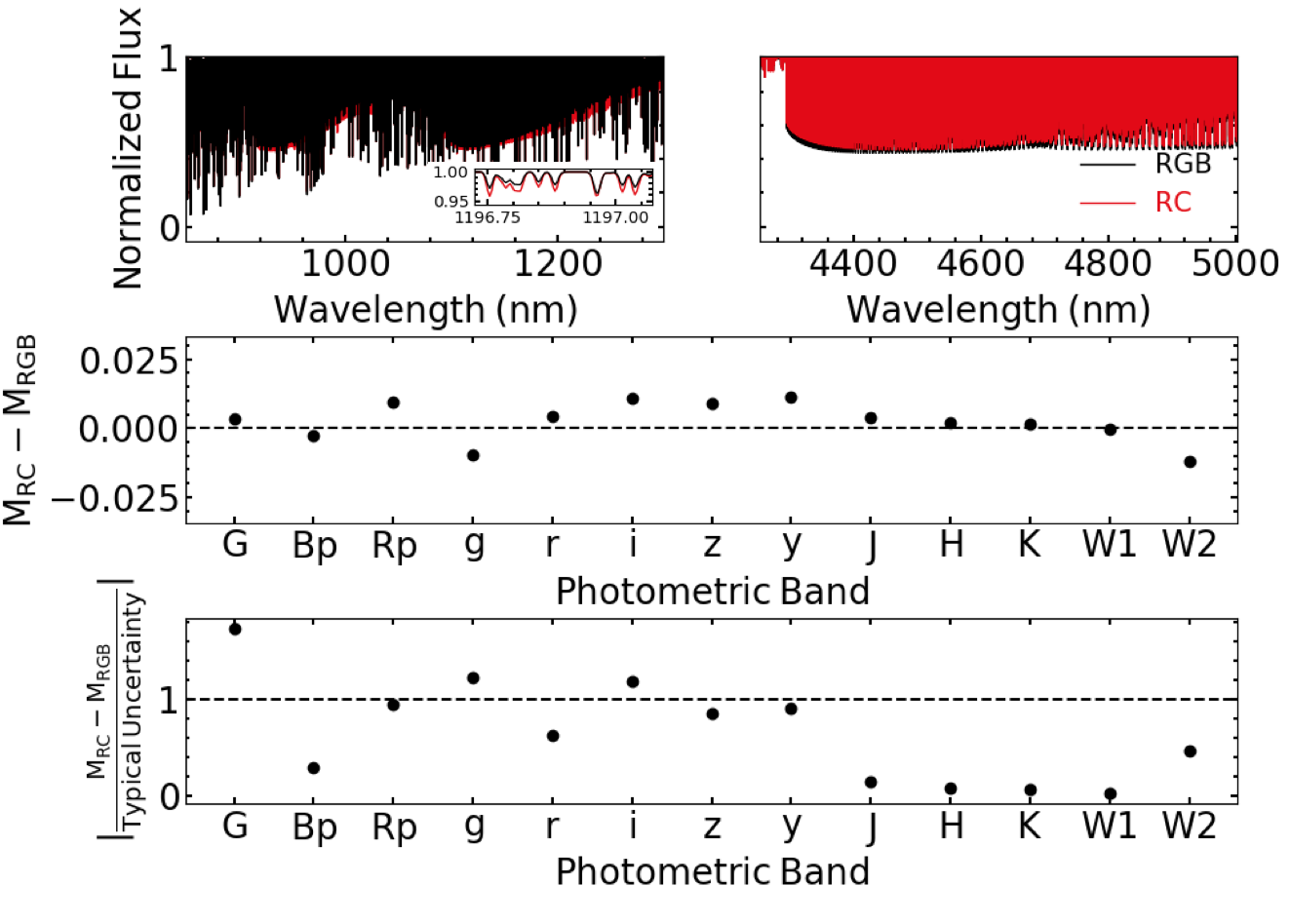}
    \caption{In the top panel we compare the synthetic spectra of an RGB and RC star. Specifically, we show two CN molecular bands which impact the y and J photometry (top left) and a CO band which impacts the W2 photometry (top right). In the top left panel we also show a zoomed-in spectrum. Both stars are synthesized with  \teff\ = 4875 K, \logg\ = 2.32 dex and Z=0.004. The RGB star has [C/N] = 0 while the RC star has [C/N]=-0.79. The increase in the N abundance causes more absorption in the CN bands which increase the magnitudes for the z, y and J passbands. On the other hand, the decrease in C causes less absorption by CO molecules which decreases the W2 magnitude. In the middle panel, we show the impacts of a change in [C/N] of -0.79 on each photometric passbands used in this work. These values are calculated using the synthetic spectra shown above. In the lower panel we show how the change in magnitudes shown in the middle panel relate to the typical uncertainty in that passband. Most bands have the [C/N] signal on the order of the uncertainties making it marginally detectable. }
    \label{fig:synth}
\end{figure*}

It is reasonable to assume that if the information to accurately select RC stars is in the spectra, then the same information is the SED just with a much weaker signal. In Figure \ref{fig:synth}, we show the theoretical differences in magnitudes between an RGB and RC star. These values are calculated using {\small PYPHOT}\footnote{https://mfouesneau.github.io/docs/pyphot} and the synthetic spectra shown in the upper panel of Figure \ref{fig:synth}. The spectra are synthesized in 1D local thermodynamic equilibrium (LTE) using ATLAS12 and SYNTHE maintained by R.Kurucz \citep[and references therein]{Kurucz1970,Kurucz1981,Kurucz1993, Kurucz2005, Kurucz2013, Kurucz2017}. We use a resolution of R$\sim$ 300,000 and the line list provided by R. Kurucz\footnote{http://kurucz.harvard.edu}. Both spectra are calculated at \teff\ = 4875 K, \logg\ = 2.32 dex and Z =0.004. The RGB spectra has [C/N] = 0 while the RC spectra has [C/N] = -0.79. These \teff, \logg\ and [C/N] values are the theoretical values of a RC star with mass of 1.00 $\rm{M_{\odot}}$ and Z = 0.004 \citep{Lagarde2012}. The change in the [C/N] ratio is caused by an increase in the nitrogen abundance as well as a decrease in the carbon abundance. In addition to this change, the CNO cycle causes a slight decrease in the oxygen abundance. This decrease in the oxygen abundance will decrease the abundance of CO molecules which releases carbon atoms. Combined with the increase in nitrogen, this leads to an increase in the abundance of CN molecules \citep{Ting2018b}. Therefore, specific photometric bands will experience higher magnitudes due to more absorption from CN molecules. Additionally, the magnitude for other bands will increase due to less absorption from CO molecules. The decrease in oxygen is a much smaller effect and is not taken into account in Figure \ref{fig:synth}. Therefore, our estimations in the change of magnitudes between RGB and RC stars are likely slightly underestimated. As shown in the lower panel of Figure \ref{fig:synth}, the difference in magnitude is typically on the order of the uncertainties (0.002, 0.010, 0.010, 0.008, 0.007, 0.009, 0.011, 0.012, 0.026, 0.027, 0.026, 0.026, 0.026 mag for G, BP, RP, g, r, i, z, y, J, H, K, W1 and W2, respectively). This makes it challenging to pick up this signal. Therefore, we use a neural network which is flexible and sensitive enough to detect the photometric signal of a [C/N] change  in RC stars. We confirm the network harnesses this information to perform the inference in Section \ref{sec:int_p}.

In order to train the network and evaluate the method we require a set of data with known spectroscopic and asteroseismic parameters. We describe the data set we use to train and validate the network in Section \ref{sec:traindata}. The validation set is used to evaluate the accuracy of the neural network. It is comprised of data with known spectroscopic and asteroseismic parameters that were never exposed to the network for the training. Therefore, it can be used determine the networks precision on the data we wish in infer parameters for. This set is also described in Section \ref{sec:traindata}. In Section \ref{sec:data} we describe the photometric data that for which we infer \teff, \logg, \ps\ and \fsep\  in order to select RC stars.

\subsection{Training and Validation Data} \label{sec:traindata}
The accuracy of neural networks highly depends on the size of the training set. Specifically, neural networks typically outperform other machine learning algorithms when the training data set is large. Therefore, we choose to use the large catalog of spectroscopically derived \ps\ and \fsep\ parameters from \citet{Ting2018} rather than smaller catalogs of asteroseismically derived parameters.  The catalog from \citet{Ting2018} is derived using a data-driven neural network which finds a mapping between the normalized spectral fluxes to the asteroseismic parameters \ps\ and \fsep.  It then infers on LAMOST DR3 \citep{Xiang2017} spectra that are within a 3D convex hull\footnote{The convex hull is the smallest polygon that contains all of the data. For more information on convex hulls see \citet{Ting2016}} of  \teff, \logg\  and [Fe/H] values that are within the parameter range of the training set used \citep[see Figure 3 in][]{Ting2018}. For our training set we include only high quality \ps\ and \fsep\ measurements from this catalog, requiring the spectra to have signal-to-noise per pixel S/$\rm{N_{pix}}$ >75 $\rm{pixel^{-1}}$. With this quality cut, the training sample has a contamination rate of $\sim$3\% \citep{Ting2018}. We train the \ps\ and \fsep\ network on a subset of 30,000 stars from this sample. The rest serves as a validation set. We include data with S/$\rm{N_{pix}}$ < 75 $\rm{pixel^{-1}}$ in the validation set of 100,000 of stars to determine the accuracy of the inference on lower S/N data. Since the LAMOST sample has a magnitude limit that corresponds to a RC distance that is greater than that given by the AllWISE magnitude limit, this test sample spans the entire distance range of our data set for inference. Therefore, the validation sample is representative of our final sample.  

In addition to this training set we require another training set to determine photometric \teff\ and \logg\ outside of the convex hull in order to separate the dwarf stars from the giant stars in the data. We  adopt the \teff\ and \logg\ values from the LAMOST DR3 stellar catalog as our training and validation sets for these separate networks (see Section \ref{sec:selectgiant}). Here, we require the training set of 200,000 to have LAMOST g-band S/N > 50. Again, our validation set of 900,000 stars does not have a S/N cut in order to cover the same magnitude range as our final sample.

With both training sets we perform a sky crossmatch with \gaia\ DR2. We then use the provided \gaia\ DR2 crossmatches \citep{Marrese2019} to obtain the photometry from AllWISE, 2MASS, and Pan-STARRS1.

\subsection{SEDs \& Parallaxes for Inference} \label{sec:data}
For the inference, we only require photometry and parallax. Therefore, our sample is much larger than spectroscopic samples given that spectra require longer exposures because of the need for more flux than imaging. We can also take advantage of the hundreds of millions of stars that already have photometry that spans from the near-UV to the mid-infrared from large surveys.

In this work we make use of data from \gaia\ DR2 \citep{Gaia2018}, Pan-STARRS1 \citep{Chambers2016}, 2MASS \citep{Skrutskie2006} and AllWISE \citep{Wright2010,Mainzer2011} photometric catalogs, along with \gaia\ DR2 parallaxes. We include all the pass bands from \gaia\ (G, BP, RP), Pan-STARRS (g,r,i,z,y)  and 2MASS (J, H, K). We use only the two bluest AllWISE bands, W1 and W2. The two other bands, W3 and W4, are shallower, have lower spatial resolution and do not contain much additional information about the SED of RC stars. We also make use of the provided \gaia\ DR2 crossmatches with Pan-STARRS1, 2MASS, and AllWISE which take into account the motions of the targets and the varying epochs of the different surveys \citep{Marrese2019}.

We perform multiple quality cuts to ensure we only use accurate photometry and parallaxes:

\begin{enumerate}
    \item 1.0+0.015$\rm{(BP-RP)^2}$ <phot\_bp\_rp\_excess\_factor< 1.3 + 0.060$\rm{(BP-RP)^2}$ \citep{Arenou2018, Evans2018}
    \item RUWE (renormalized unit weight error) <1.4 \citep{Lindegren2018b}
    \item ``A" quality photometry from 2MASS and AllWISE
    \item ($\sigma_G$\footnote{$\sigma_G = \sqrt{\left(\frac{\sigma_I}{I}\right)^2 + \sigma_{ZP}^2}$ where $\frac{\sigma_I}{I}$ is the mean G band error over the flux and $\sigma_{ZP}$ is the zero-point error which for \gaia\ DR2 is 0.0018 mag} $-0.0018)/G <0.0002$ to remove nearby stars that are near saturation in \gaia\ DR2

\end{enumerate}

We do not apply a quality cut to the Pan-STARRS (PS1) catalog because the Image Processing Pipeline (IPP) \citep{Magnier2016} achieves relative photometric calibration better than 1\% for much of the sky \citep{Schlafly2012}.

Since neural networks perform poorly on data that is dissimilar from what it has been trained on $-$i.e., the network is not portable and cannot extrapolate, we create a convex hull in the photometric and parallax space based on the validation set and do not infer on any data that does not fall within that convex hull. To create the convex hull we randomly select 10,000 stars from our training set and use the BP, g, K, W1 and W2 bands along with the parallax. As the convex hull algorithm can only handle up to 6 dimensions we choose a subset of the photometric bands to use. Bp, g, K, W1 and W2 are selected because we want to span the entire wavelength range while also prioritizing bands important for the \ps\ inference (see Figure \ref{fig:synth}) and infrared bands that will be least impacted by extinction. Although, we find that changes in the band selection has an  insignificant impact on the final red clump sample.

\section{Method}
\label{sec:method}
From synthetic photometry, we know that the signal of the evolved [C/N] ratio that we hope to use to infer the \ps\ is much smaller than the effects of \teff\ and extinction on the photometry. It is also on the order of the noise in the photometry (see Figure \ref{fig:synth}). Given these constraints and access to a large training set,  we find a neural network with probabilistic inference to be an effective tool to infer the parameters from the photometry. In this work, we infer the \ps, \fsep, \teff, and \logg\ of stars from 13 photometric bands and \gaia\ DR2 parallaxes using Mixture Density Networks \citep[MDNs;][]{Bishop94}. We use 6 separate networks, 2 each for \teff\ and \logg\ for all stars and for just giant stars and 2 for \ps\ and \fsep\ of giant stars. We defer the simultaneous inference of multiple parameters with a single MDN which will better account for covariances to a future study. 

\subsection{Mixture Density Network} \label{sec:MDN}
 A MDN is a neural network where the outputs parametrize a Gaussian mixture model (GMM):
\begin{equation}
    p(\theta|x) = \Sigma_{j=1}^{n}\varpi_j(x)\mathcal{N}(\mu_j(x),\sigma_j(x))
\end{equation}
where one can choose the number of components ($n$). Unlike a typical GMM, the weights ($\varpi(x)$), and widths ($\sigma(x)$) along with the means ($\mu(x)$) of each component are a function of the input parameters ($x$) which in our case is the 13-band SED plus parallax.  In other words, given an SED and parallax ($x$), we expect the output parameter ($\theta$), in this case \teff, \logg, \ps, or \fsep, can span a distribution, and we try to model that distribution using the training data, essentially doing a conditional density estimation of the training set. Having estimated such density during the training, when the network is exposed to new input data for the inference, we can then read off the probability density of the output parameter (\teff, \logg, \ps, or \fsep). Thus the outputs of our networks ($\varpi(x)$,$\sigma(x)$, and $\mu(x)$) parameterize a GMM which acts as a probability distribution function (PDF) and we use the negative log likelihood as the loss function. In other words, we train the network to maximize the sum of the log likelihoods of the output PDFs given the training data. This method differs from \citet{Ting2018} in that we are essentially doing a density estimation of the training set with a neural network while \citet{Ting2018} performs non-linear regression with their neural network.

The advantage of using a MDN over a standard neural network is that the GMM output is probabilistic instead of single best-estimate and therefore gives as the ability to estimate uncertainties for our inferred parameters. However, these uncertainty estimates assumes there is no uncertainty in the input parameters (SED and parallax). Hence, we do not perform any input uncertainty propagation. In practice, we can also integrate the input parameter uncertainties to get the full posterior. For the purpose of this study, in which we are more focused on classifying RC stars, our reported uncertainties do not impact the classification and therefore we do not calculate the full posterior. Therefore, our uncertainties are a reflection of the training set density distribution. However, since the training set density distribution has noise, the uncertainty estimate does capture some of the input noise (see Section \ref{sec:int_t_g}).  An additional benefit to the MDN is that we can use a two-component GMM output for the \ps\ network in order to encapsulate the bi-modality of the posterior distribution of \ps. Otherwise we use a one-component network for the \teff, \logg, and \fsep\ inference. When we use a one component model, the mean ($\mu(x)$) and width ($\sigma(x)$) maps directly to the inferred values and uncertainty. However for the \ps\ network when we use two-component, the inferred value is the weighted mean of the means ($\Sigma_{j=1}^2\omega_j(x)\mu_j(x)$) and the adopted uncertainty is the weighted mean of the widths ($\Sigma_{j=1}^2\omega_j(x)\sigma_j(x)$) Besides the number of components of the output GMMs, each of the MDNs use the same architecture. We use four fully connected layers with 32, 24, 16 and 8 nodes to map from the 14 input parameters (SED + parallax) to the $3\times n$ outputs that parametrize the GMM ($\omega(x)$,$\sigma(x)$ and $\mu(x)$). We apply a sigmoid activation function\footnote{The choice in activation function is rather arbitrary here given the low-dimensionality of the inputs (14) and outputs ($3\times n$ where $n$ is 1 or 2)} to each node to capture the non-linear mapping from the inputs to the outputs. 


\subsection{Inferring \teff\ and \logg\ and Selecting Giant Stars} \label{sec:selectgiant}


\begin{figure*}
    \centering
    \includegraphics[width=\linewidth]{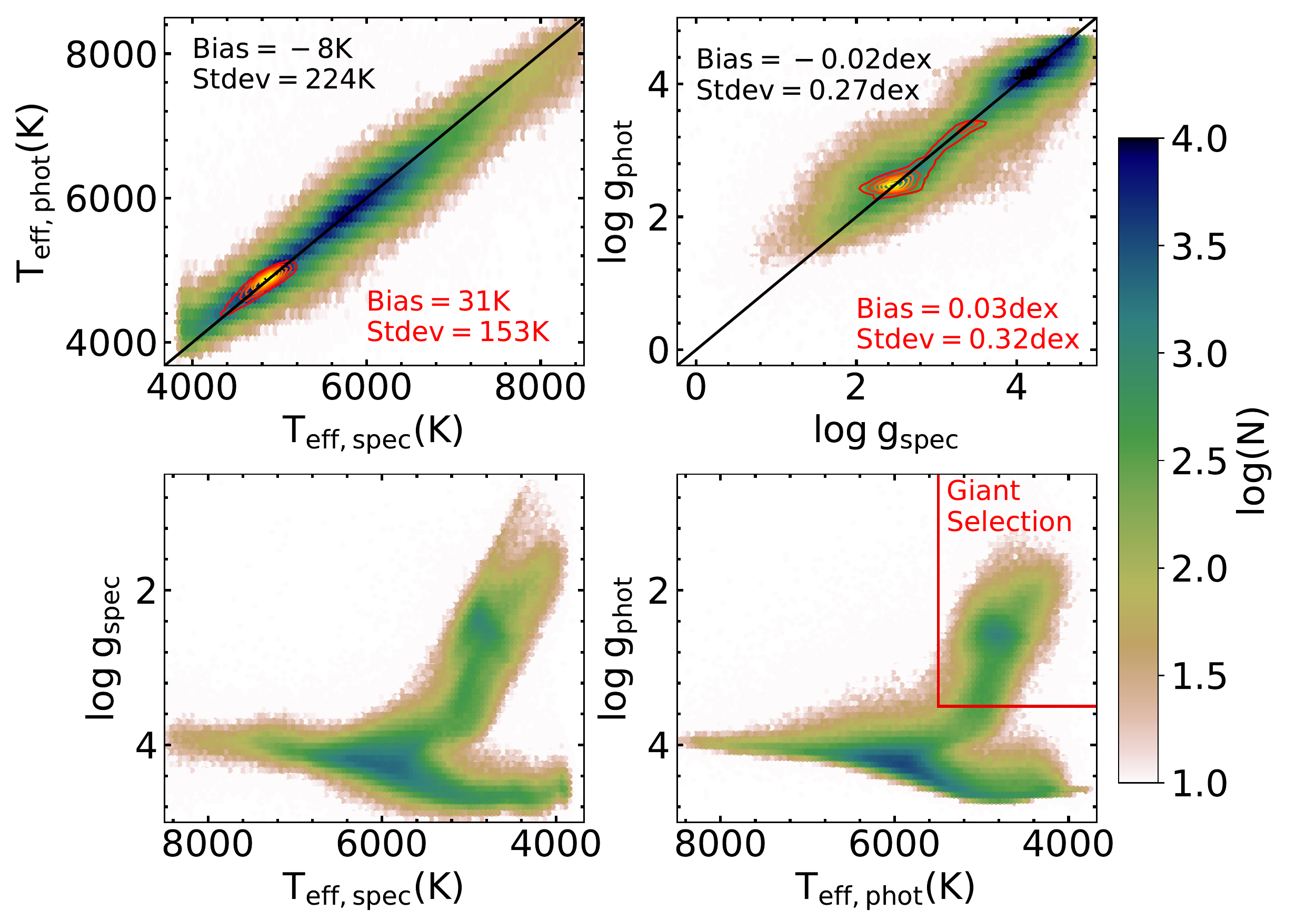}
    \caption{Comparing the stellar parameters (\teff\ and \logg) we derive from the photometry compared to the spectroscopically derived values from LAMOST using the validation set which was not used to train the network. The top panels show directly how the inferred values compare to the spectroscopic values for the entire validation set. The results from the first \teff\ and \logg\ networks which all of the data go through are shown in the beige to blue density plot. The bias and standard deviation for this inference are shown in black. In the red to yellow contours we show our refined inference which uses networks trained on only spectroscopic giant stars and applied to photometrically selected giant stars. The inference improves when we retrain the network with a smaller parameter range. We show the statistics for this inference in red. In the bottom panel, we show the LAMOST Kiel diagram (\teff-\logg\ ``Hertzsprung-Russell" diagram) for the validation sample (left) and  the inferred photometric Kiel diagram (right). The main sequence and giant branch are preserved in the photometric diagram. Therefore we can confidently select giant stars using our inferred \teff\ and \logg. The red box in the photometric Kiel diagram (right) shows the selection of giant stars which we determine \teff\ and \logg\ using the refined networks. }
    \label{fig:teff_logg}
\end{figure*}

We first infer the \logg\ and \teff\ of all of the stars in the sample in order to select the giant stars. To do this, we train two MDNs using a random sample of 200,000 stars from the LAMOST DR3 catalog which pass the quality cuts (see Section \ref{sec:traindata}). Both of these networks have one mixture component as the output so the inferred value is the mean of the Gaussian and the uncertainty is the width.  We show a comparison of the derived values with the LAMOST results for 900,000 validation stars in Figure \ref{fig:teff_logg}. In order to quantitatively describe the accuracy and precision of the inferred parameters, we report the bias ($\langle \rm{LAMOST}-\mu(\textit{x})\rangle$), the standard deviation ($\langle (\rm{LAMOST}-\mu(\textit{x}))^2\rangle$) and the typical uncertainty ($\langle\sigma(x)\rangle$). However, as we noted in Section \ref{sec:MDN}, the uncertainty estimate does not fully take into account the uncertainty on the input data. Therefore, it is likely that our uncertainties are underestimated, especially given that the network is trained on higher S/N data. The standard deviation from external validation is a more accurate estimate of the precision. Nonetheless, as we will see, the standard deviation is typically on the order of the uncertainty estimate $\langle\sigma(x)\rangle \approx \langle (\rm{LAMOST}-\mu(\textit{x}))^2\rangle$. This shows that, since the training input is noisy, the adopted uncertainty does capture some of the input uncertainty. The bias for photometric \teff\ values is -8 K with a standard deviation of 224 K. The typical uncertainty on \teff\ is 200 K which is reasonable given that the typical uncertainty for \teff\ in the LAMOST catalog is 168 K. We also find a bias for photometric \logg\ values of -0.02 dex with a standard deviation of 0.27 dex. The typical \logg\ uncertainty is 0.23 dex which is again consistent with the typical uncertainty for \logg\ in the LAMOST catalog is 0.24 dex. The lack of bias in the \teff\ and \logg\  is expected given that we are using a data-driven method and the parameters are tied to the scale of the training set. After inferring on all of the data that passes the quality cuts (see Section \ref{sec:data}), we select stars with inferred \logg\ < 3.5 dex and 2500 K < \teff\ <5500 K as giant stars (see Figure \ref{fig:teff_logg}).

Once the giant sample is created we refine the \teff\ and \logg\ inference by using a network trained on the spectroscopic parameters of just giant stars. This network is trained on a subset of 200,000 giants stars selected using the method described above. We use the initial photometrically inferred parameters to make the giant selection for the training set to ensure our training set will be similar to the one we will use for the inference, which will only have photometrically inferred parameters.  Again, we use one mixture component as the output so the inferred value is the mean of the Gaussian and the uncertainty is the width. The red contours in Figure \ref{fig:teff_logg}, demonstrate how the inference has been refined. These new photometric \teff\ values have a bias of 31 K  with a standard deviation of 153 K. The typical \teff\ uncertainty is 97 K. The bias of photometric \logg\ values is 0.03 dex with a standard deviation of 0.32 dex. The typical uncertainty is 0.22 dex. The typical uncertainties for \teff\ and \logg\ reported here (97 K and 0.22 dex) are smaller than the typical uncertainty in the LAMOST catalog (168 K and 0.24 dex). As explained, this is likely because  our uncertainties are underestimated by not taking into account the uncertainty on the input data. In addition, the training set has higher S/N than the validation set. The external validation gives us a more robust estimate of the precision of the inference.

\begin{figure*}
    \centering
    \includegraphics[width=\linewidth]{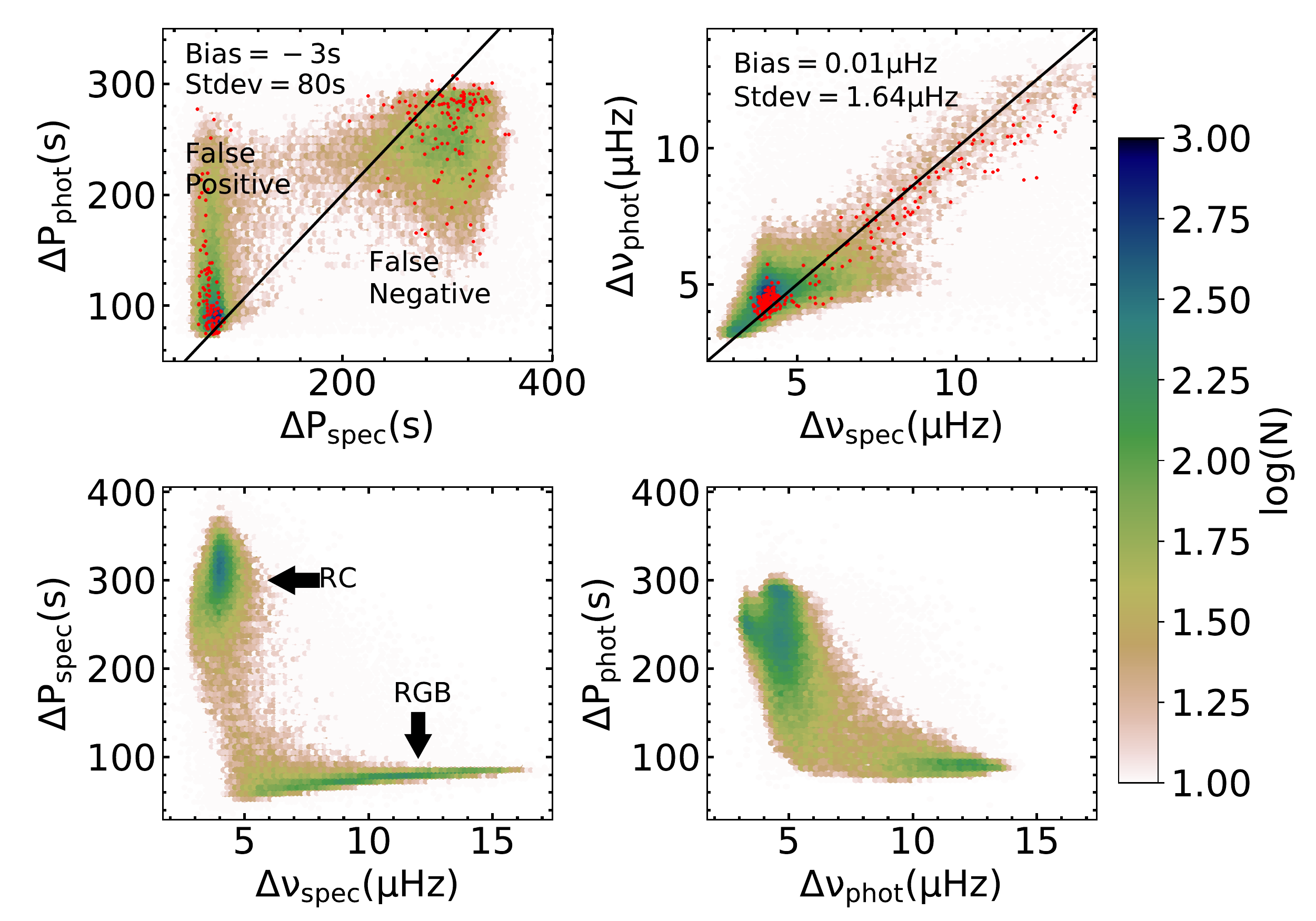}
    \caption{The asteroseismic parameters (\ps\ and \fsep) we derive for giant stars compared to the spectroscopically derived values from LAMOST spectra \citep{Ting2018}. The red points in the top panels compare our derived photometric values with asteroseismc parameters from \citet{Vrard2016}. The \ps\ is the most effective parameter for selecting RC stars since they have \ps\ values more than 100s larger than RGB \ps\ values. From the bottom left panel, we can see that RC stars have \fsep\ $\lesssim$ 5 \michz\ and \ps\ $\gtrsim$ 200s. While, RGB stars have \fsep $\gtrsim$ 5 \michz\ and \ps\ $\lesssim$ 100s. From the top right panel, we can see we are effective at picking up all of the RC stars as there are few false negatives. However, we do suffer from contamination with false positives. Combining the \ps\ with \teff, \logg\ and \fsep\ can help limit this contamination. On the bottom left we show the asteroseismic parameters (\ps\ and \fsep) derived from LAMOST spectra in \citet{Ting2018}. On the bottom right we show the derived asteroseismic parameters from photometry. We can pick out the RC with \ps\ > 200 s from the photometrically inferred asteroseismic parameters.  }
    \label{fig:ps_fs}
\end{figure*}

\subsection{Inferring Asteroseismic Parameters and Selecting Red Clump Stars}

We make the selection of RC stars using inferred \logg, \teff, \fsep\ and \ps. For each of these parameters we train another MDN. For \fsep\ and \ps, we train on a subset of 30,000 stars from the sample presented in \citet{Ting2018}. For the \fsep\ the network also has one mixture component as the output and we infer using the same method as \teff\ and \logg. The distribution of \ps\ is bi-modal (see Figure \ref{fig:ps_fs}). Using two mixture components helps the network to learn to reproduce this bi-modality, especially when the input is noisy and the posterior is likely bimodal. Therefore, the output of the \ps\ network is a two component Gaussian mixture model. We define the inferred \ps\ to be the weighted mean of the means of the two components and the uncertainty is the weighted mean of the widths of the two components. As we are trying to infer the \ps\ based on the signal of a change [C/N] in the photometry, which is roughly on the order of the uncertainty (see Figure \ref{fig:synth}), we know that the network will not be perfect. However, the inference result is encouraging (see Figure \ref{fig:ps_fs}). The bias for the photometric \fsep\ is only 0.01 \michz\  with a standard deviation of 1.64 \michz. The typical uncertainty is 1.51 \michz. For \ps, the bias is -4 s with a standard deviation of 80 s. The typical uncertainty is 37 s. Although this uncertainty is not negligible, it is much smaller than the expected difference in \ps\ between RGB and RC stars ($\sim$ 200 s). We also compare our photometrically derived values to asteroseismically derived values from \citet{Vrard2016}. We find a bias of 0.27 \michz\ and a standard deviation of 1.28 \michz\ for \fsep\ and a bias of 6 s and standard deviation 61 s for \ps. It is interesting to note that the biases are slightly higher for the asteroseismic comparison while the standard deviations are slightly smaller. The bias is likely higher because it is compounding with the spectroscopically derived bias. The standard deviation likely shrinks because the asteroseismic sample is more nearby and therefore has less noisy photometry, which improves the performance of the network.

\begin{figure*}
    \centering
    \includegraphics[width=\linewidth]{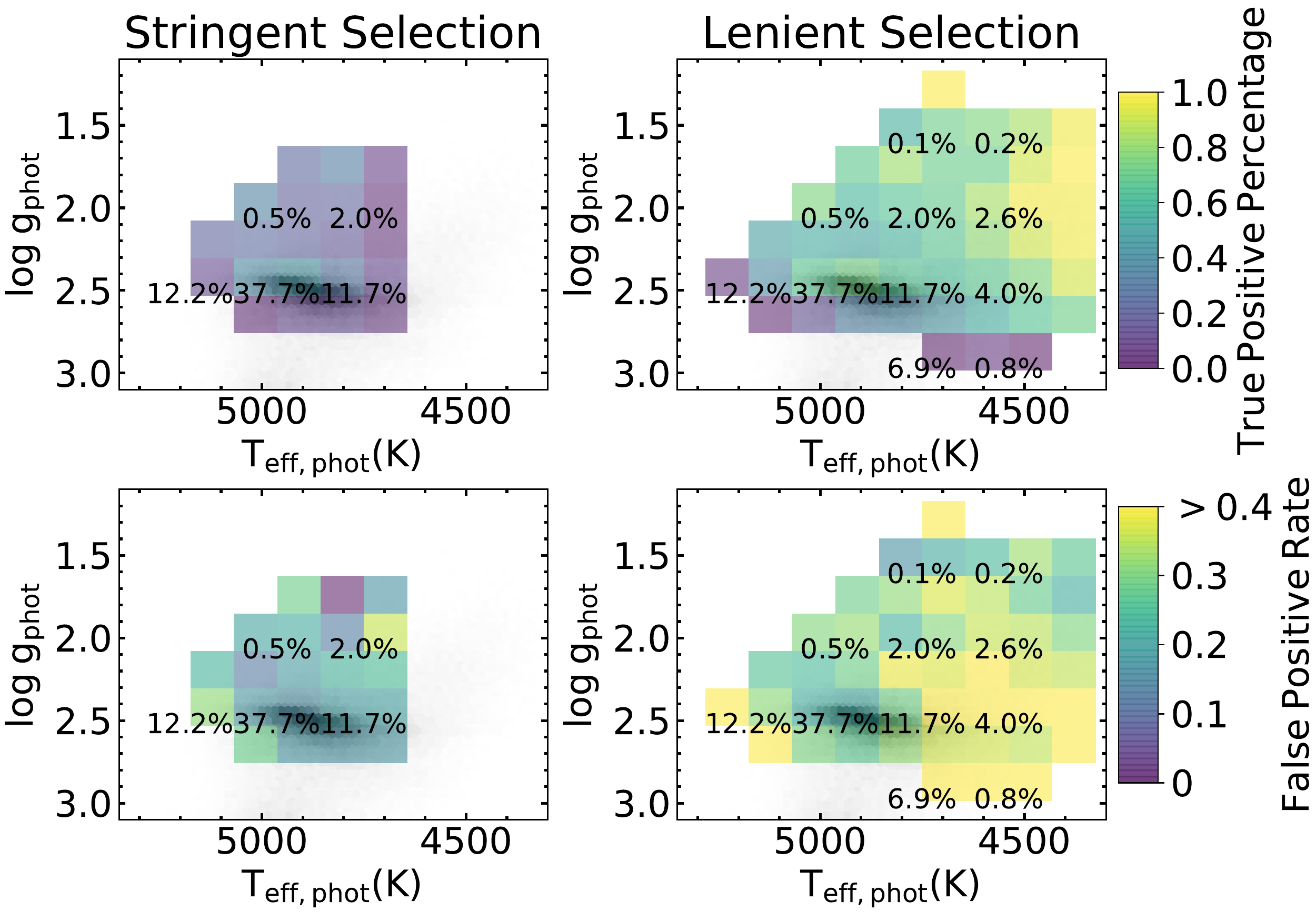}
    \caption{The true positive percentage (upper panels) and the contamination rate (lower panels) as a function of the photometrically derived \teff\ and \logg\ for both of our recommended samples. These are evaluated using the spectroscopic sample for which we know the ``ground truth". The black text shows the percentage of true spectroscopic RC stars in each sample that are in the adjacent four bins. We define the contamination rate to be the number of false positives divided by the total number of the stars selected per box. The true positive percentage is defined as the number of stars successfully identified as RC divided by the number of true RC stars in the selected box. The panels on the left correspond to our stringent one selection which is \ps\ >265 s, \fsep\ < 5.25 \michz\ and \teff\ >4700 K. The panels on the right correspond to a more lenient selection of \ps\ >225 s and \fsep\ < 5.5 \michz. When we soften our selection criteria the true positive percentage increases. However, so does the contamination rate. Integrating over all of the bins, the tier 1 (stringent) selection leads to an overall contamination rate of $\sim$ 20\% and overall true positive percentage of $\sim$ 25\%. While the tier 2 (lenient) selection leads to an overall contamination rate of $\sim$ 33\% and overall true positive percentage of $\sim$ 94\%.}
    \label{fig:binning}
\end{figure*}

To find the ideal RC selection criteria, we have to balance completeness with the contamination. In general, we want to maximize the completeness while minimizing the contamination. However, there may be science cases where one may wish to sacrifice completeness for a more pristine sample or vice versa. In this work, we quantify completeness by the true positive rate, which is defined as the percentage of true spectroscopic RC stars in the validation sample that are successfully identified as RC stars  (1-false negative rate, see Figure \ref{fig:ps_fs}). We quantify the contamination rate as the percentage of stars which are selected as RC stars that are not true RC stars (false positive rate, see Figure \ref{fig:ps_fs}). We expect the contamination rate and true positive percentage to be a function of \teff\ and \logg\ along with \fsep\ and \ps\ because the density of true RC versus RGB stars varies as a function of \teff\ and \logg. Therefore, we decide to bin the data in \teff, \logg, \fsep\ and \ps\ in order to find the ideal selection criteria.  Using our validation set, we bin the data using the photometrically inferred parameters and calculate the contamination and true positive rate within each bin. Although this is done in four dimensions (\teff, \logg, \fsep\ and \ps), we show flattened two dimensional examples in Figure \ref{fig:binning}. It is interesting to note that the contamination rate for the tier 2 (lenient) sample (top right panel) may be artificially low at the lowest \teff\ and \logg\ values. This is likely because the tip of the red giant branch stars that are located there would have a [C/N] most similar to the RC values. However, the fraction of stars in the region is limited. We bin each parameter into 20 bins so that in total we have 160,000 bins. This results in bin sizes of roughly 50K for \teff, 0.10 dex for \logg, 1 \michz\ for \fsep\ and 20 s for \ps. It is important to note that we bin using the photometrically derived parameters so that the results are applicable to our final sample which do not always have spectroscopically derived values. We include bins with a low contamination while also having a significant percentage of the true RC stars. Further information about the final sample can be found in Section \ref{sec:sample}.

\subsection{Deriving Distances} \label{sec:dist}
Once we have the RC sample, we infer the heliocentric distances using the AllWISE W1 band similar to \citet{Ting2019}. First, we perform an extinction correction by finding a linear relationship between the photometric \teff\ and the G$-$W1 color for less extincted stars. We then use this along with $\rm{A_G/A_{W1}}$=16 \citep{Hawkins2017} to find the extinction in the W1 magnitude and correct for it. Next, we again use the less extincted stars with \gaia\ parallax uncertainties < 5\% to derive a  linear relationship between the inferred \teff\ and the intrinsic $\rm{M_{W1}}$. In order for our absolute W1 magnitudes to be consistent with the calibrated absolute W1 magnitude for RC stars ($\rm{M_{W1}}$ =-1.68$\pm$0.02 mag) from \citet{Hawkins2017}, which is consistent with \citet{Ruiz-Dern2018}, we use a parallax offset of 0.04 mas. This parallax offset is roughly consistent with those found using other methods \citep{Lindegren2018,Schonrich2019}. If we do not include the parallax offset we find an absolute W1 magnitude of -1.99 mag. Finally, we derive the distance using the distance modulus with the inferred $\rm{M_{W1}}$ and the extinction corrected W1 magnitudes. In principal, a Bayesian approach would be more appropriate for determining the distances \citep[e.g.,][]{Hawkins2017,Anders2019}. However, the main focus of this work is to derive a clean sample of RC stars and we will refine the distances in future work. Comparing to \gaia\ parallaxes with uncertainties <5\% and assuming the \gaia\ distance as ground truth, we find our distances have uncertainties $\sim$ 9\%. This is consistent with reported theoretical precision estimates given the intrinsic dispersion in the absolute magnitude of RC stars \citep{Bovy2014,Hawkins2017}.

\section{The Red Clump Sample}
\label{sec:sample}

 \begin{figure}
     \centering
     \includegraphics[width=\columnwidth]{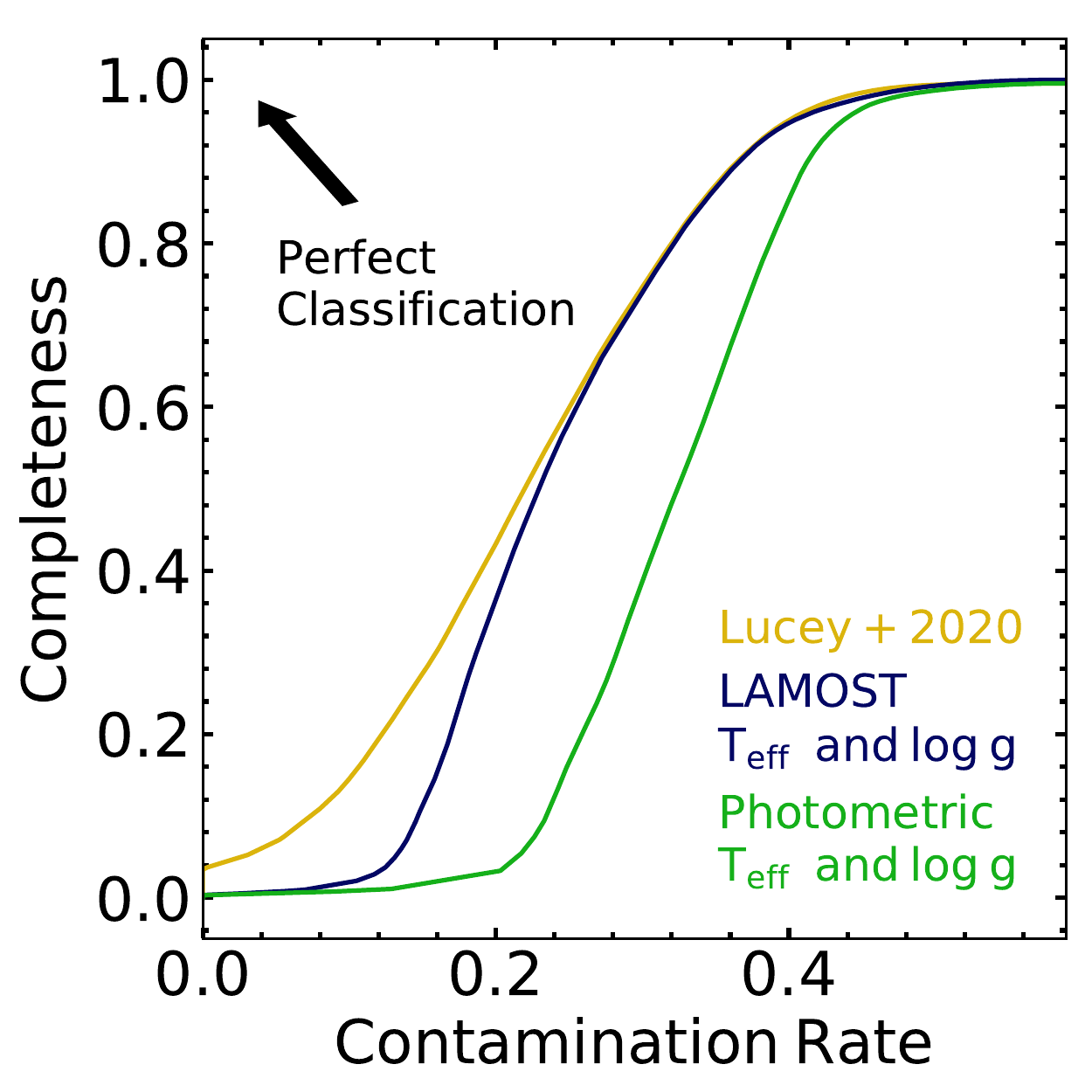}
     \caption{The receiver operating characteristic (ROC) curve of the selection method compared to other methods. The ideal classification method gives 100\% completion with no contamination. Therefore, the closer the ROC curves gets to the top left corner of the plot, the more effective the classification method. The lines are calculated by cumulatively summing bins in the order of lowest contamination rate (see Figure \ref{fig:binning}). The result for our method where we bin in all four photometrically derived parameters (\teff, \logg, \fsep, and \ps) and select from them is shown in yellow. The result of only selecting in the photometric \teff\ and \logg\ space is shown in green. The results of making the same selection but using LAMOST spectroscopically (R$\sim$ 1800) derived \teff and \logg\ is shown in blue. This shows our method can obtain a lower contamination rate than the other two standard methods.}
     \label{fig:roc}
 \end{figure}

Having inferred four parameters (\teff, \logg, \fsep\ and \ps) for each giant stars in our sample, we must now determine how we will use these parameters to determine our RC sample. Different selection criteria will result in varying completeness and contamination rates, given that we know the true positive percentage and contamination rate vary as a function of \teff, \logg, \fsep\ and \ps\ (see Figure \ref{fig:binning}). Ideally, we would like to find a selection that maximizes completeness while minimizing contamination. In this work, we prioritize a low contamination over a complete sample. However, different science cases may require different selection criteria. Various selections can make retracing the selection function more difficult. Therefore, along with the recommended samples, we provide  all of the inferred parameters for the entire giants catalog  and encourage the reader to choose their own selection criteria. 

To choose which bins from Figure \ref{fig:binning} we will use in the final selection, we perform a cumulative summation sorted by the contamination rate. The results of the cumulative summation are shown in Figure \ref{fig:roc}. Our method is able to achieve a lower contamination rate than when selecting just using spectroscopically (at low resolution, R$\sim$2,000)  or photometrically \teff\ and \logg. We note that we do not include using [C/N] information from spectra in our comparison. Using the [C/N] from spectra would result in a lower contamination rate for the spectroscopic sample \citep{Hawkins2018,Ting2018}. However, in Figure \ref{fig:roc}, we are comparing to more ``classical" methods which are still frequently used today \citep[e.g.,][]{Wegg2019,Chan2019}. It is also important to note that completeness is evaluated based on the same sample which is confined by the spectroscopic sample to brighter magnitudes. Therefore, a given completeness fraction corresponds to a much greater number of stars for the photometric sample than the spectroscopic sample given that the photometric sample contains much more data. However,  these results are also slightly idealized. The selection method used to create Figure \ref{fig:roc} selects parameter bins in order of the lowest contamination rate rather than selecting adjacent bins as we do for our final selection. 

For ease of selection and to better model the selection function in the future, for our recommended catalogs we choose to only use adjacent bins with simple parameter boundaries. We provide two RC samples with different contamination rates and true positive percentages. Again, we provide the inferred parameters for all of the giant stars and encourage the readers to make their own selection. However, we will mark the stars in our two RC samples as ``tier 1" and ``tier 2" in the catalog. For our smaller, less contaminated sample (tier 1), we achieve a contamination rate of $\sim$20 \% (see Section \ref{sec:cont} for how these contamination rates are calculate) and a true positive percentage of $\sim$25\% which results in a final sample of $\sim$ 405,000 RC stars. These results are achieved by selecting giant stars with  \teff\ > 4700 K, \fsep\ <5.25 \michz, and \ps\ > 265 s. We also select a larger sample (tier 2) ($\sim$ 2.6 million RC stars) with a contamination rate of $\sim$33\% and a true positive percentage of $\sim$ 94\%. This sample is chosen by selecting giant stars with \fsep\ <5.5 \michz, and \ps\ > 225 s. Therefore, The ``tier 1'' sample is subset of the ``tier 2" sample. As such, everything labeled as ``tier 1" in the catalog is also included in the ``tier 2" sample.

\begin{table*}
    \caption{The cuts, contamination rate, true positive percentage and total number of stars for each of our provided samples}
    \label{tab:samples}
    \begin{tabular}{ccccccc}
    Selection & \ps\ cut & \fsep\ cut& \teff\ cut & Contamination & Completeness & \# of Stars \\
    \hline \hline
     Tier 1 &  >265 s & <5.25 \michz & > 4700 K &  20\%  &  25\% & 405,000  \\
     Tier 2 & >225 s & <5.5 \michz & None &  33\% &  94\%    & 2.6 million \\
     |b|>30 degrees & >265 s & <5.25 \michz & >4700 K &  9\% &  12\%    & 15,000 \\
     \hline
    \end{tabular}
    \flushleft{Description of the final provided samples of RC stars. We also provide the inferred \teff, \logg, \fsep\ and \ps\ in the final catalog in order for the reader to be able to choose their own selection criteria for their science case. }
\end{table*}

\begin{table*}
\caption{Photometrically derived \teff, \logg, \ps\ and \fsep\ for all giant stars along with distances for selected RC stars.}
\label{tab:catalog}
\begin{tabular}{ccccccccc}
\gaia\ Source ID & RA & DEC & \teff\ & \logg\ & \ps\ &  \fsep\  & Selection & d  \\
\hline \hline
 & (deg) & (deg) &  (K)  & (dex) & (s)  & (\michz) & & (kpc)  \\
 \hline
4130510288119401088 & 252.313 & -19.93 & 5021$\pm$119 & 2.68$\pm$0.39 & 185$\pm$40 & 7.65$\pm$2.95 &   &   \\
4114024073354437376 & 255.807 & -22.352 & 4574$\pm$89 & 2.64$\pm$0.16 & 247$\pm$44 & 4.9$\pm$1.08 & tier 2 & 4.72 \\
4109961240517001856 & 261.658 & -25.704 & 3741$\pm$ 31 & 2.49$\pm$0.2 & 185$\pm$ 42 & 0.13$\pm$0.05 &   &   \\
4114024550075584768 & 255.861 & -22.332 & 4948 $\pm$ 120 & 2.28 $\pm$0.64 & 185 $\pm$ 53 & 7.12 $\pm$10.15 &   &   \\
4114024623110255232 & 255.855 & -22.32 & 4102 $\pm$ 93 & 1.69 $\pm$ 0.24 & 255 $\pm$ 34 & 2.6 $\pm$ 0.12 & tier 2 & 2.94 \\
4114027539372846848 & 255.743 & -22.316 & 4956 $\pm$ 119 & 2.99 $\pm$ 0.72 & 274 $\pm$ 37 & 10.71 $\pm$ 3.1 &   &   \\
4114027917328986240 & 255.807 & -22.281 & 5037 $\pm$ 134 & 3.23 $\pm$ 0.18 & 103 $\pm$ 29 & 12.57 $\pm$ 1.75 &   &   \\
4126036886406973952 & 255.633 & -22.265 & 4035 $\pm$ 84 & 2.45 $\pm$ 0.29 & 229 $\pm$ 33 & 2.11 $\pm$ 0.16 & tier 2 & 4.13 \\
6734964158462988160 & 277.756 & -33.984 & 4920 $\pm$ 144 & 2.39 $\pm$ 0.58 & 315 $\pm$ 26 & 4.59 $\pm$ 1.15 & tier 1 & 7.86 \\
6734962754045102848 & 277.679 & -34.036 & 4849 $\pm$ 223 & 2.86 $\pm$ 0.05 & 76 $\pm$ 4 & 9.59 $\pm$ 2.82 &   &   \\
... & ... & ... & ... & ... & ... & ... & ... & ...\\
\hline
\end{tabular}
\flushleft{A sample of the provided catalog with \teff, \logg, \ps\ and \fsep\ for all giant stars along with uncertainties. For a cautionary note about the uncertainties, see Section \ref{sec:int_t_g}. We also give the ``tier 1" and ``tier 2" RC selection as described in Section \ref{sec:sample}. The ``tier 1" is a less contaminated subset of the ``tier 2" sample. However, we only label it as ``tier 1" in the catalog for simplicity. We provide distance estimates for stars selected as RC.  }
\end{table*}

\subsection{Contamination Rate} \label{sec:cont}

\begin{figure*}
    \centering
    \includegraphics[width=\linewidth]{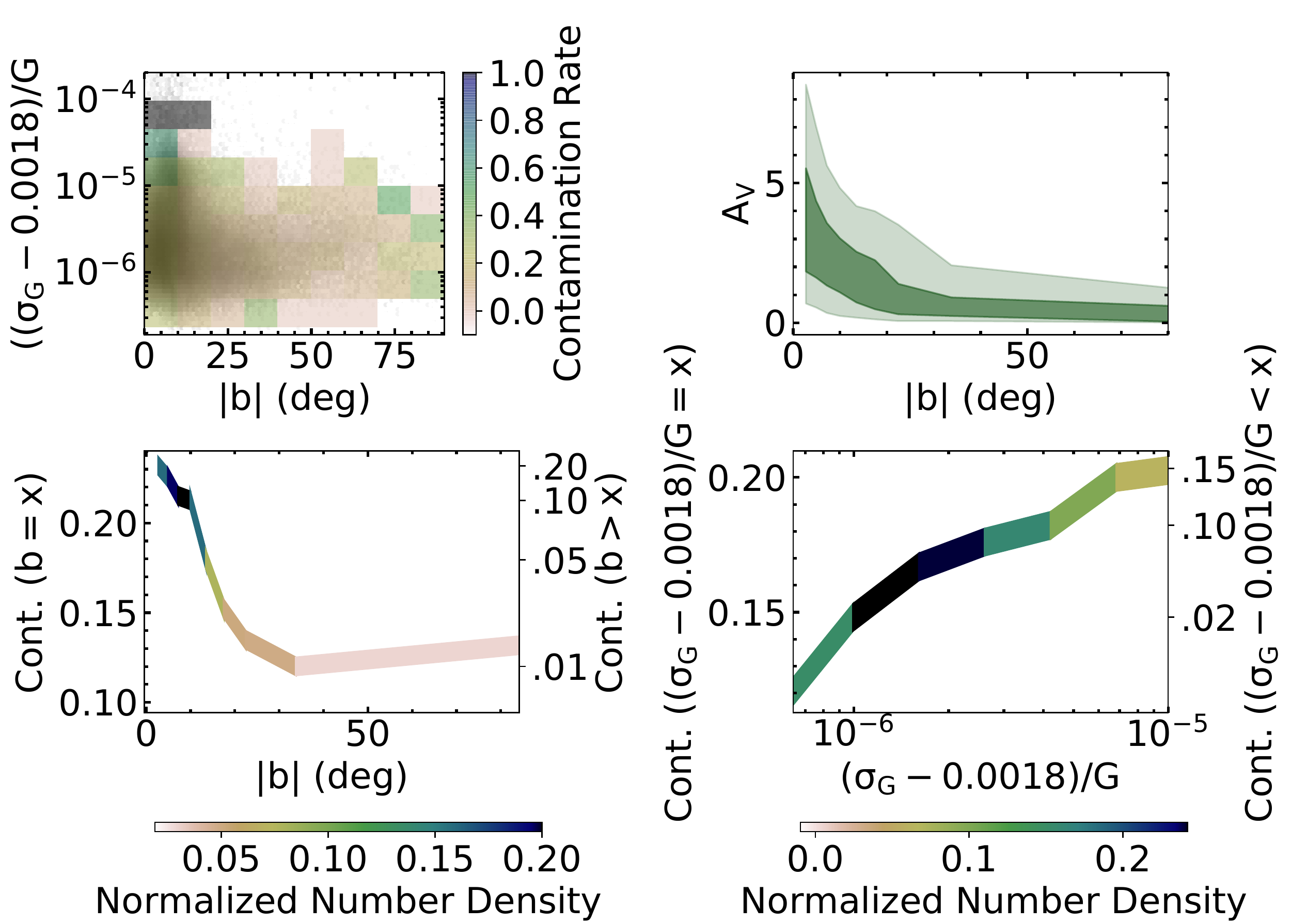}
    \caption{ The contamination rate as a function of the fractional \gaia\ G band magnitude uncertainty (($\sigma_G-0.0018)/G$), and the Galactic latitude ($b$). In the top left plot, we show the distribution of our ``tier 1" sample in G band fractional uncertainty and Galactic latitude in grey with the contamination rate in the colored boxes. In the top right plot we show the extinction ($A_V$) as a function of Galactic latitude. This shows that extinction correlates with Galactic latitude and we can therefore use Galactic latitude as a proxy for extinction when evaluating the contamination rate. In the bottom two plots, we show the 1D versions of the top left plot. For these plots, the left y-axis shows the contamination rate at x. The right y-axis shows the cumulative contamination rate for all data >x for Galactic latitude and <x for the fractional G band uncertainty. These values are calculated by summing the contamination rate multiplied by the normalized number density of the ``tier 1" sample. The lines are colored by the value of the normalized number density at that point. }
    \label{fig:cont}
\end{figure*}

In order to accurately evaluate the contamination rate and completeness of our final sample we need to determine how the contamination and completeness of our validation set behaves as a function of extinction and uncertainty in the SED. Therefore, we can account for differences between our final samples distribution in these parameters compared to our validation set. 

In Figure \ref{fig:cont}, we show the contamination rate as a function of Galactic latitude ($b$) and the fractional \gaia\ G band magnitude uncertainty (($\sigma_G-0.0018)/G$). We evaluate the fractional G band uncertainty as $(\sigma_G-0.0018)/G$ because we want to subtract off the contribution of the uncertainty in the zero-point which converts the measured flux to a magnitude using the Vega magnitude system. This conversion accounts for the majority of the uncertainty for a large fraction of the data (see bottom right panel of Figure \ref{fig:cont}), so that we have a more dynamic range of uncertainties with which to evaluate the contamination. We also test calculating the contamination rate as a function of just $\sigma_G$ and the results are similar to $\sigma_G-0.0018)/G$. In the top right panel of Figure \ref{fig:cont}, we show the extinction ($A_V$) as a function of Galactic latitude for the RC sample. We calculate the extinction ($A_V$) using the Python package {\small{DUSTMAPS}} \citep{Green2018} and the map provided by \citet{Green2019}. Ideally, we would calculate the contamination as a function of $A_V$. However, For accurate estimates of $A_V$, we require accurate distances to our objects. Given that we do not have accurate distance for the RGB stars in our validation sample, we cannot calculate $A_V$ for these stars and therefore cannot calculate the contamination rate as a function of $A_V$. However, because the extinction is highly correlated to the Galactic latitude (see Figure \ref{fig:cont}), we can use the Galactic latitude as a proxy for extinction. This correlation is expected given that in the disk, at $|b|$ <10 degrees, there are high levels of  dust, which causes extinction. However, at $|b|$ > 10 degrees, we are looking out of the plane of the Galaxy, where dust density and therefore extinction is much lower. As extinction can greatly impact the shape of an SED, we expect it to negatively impact our ability to infer the \teff, \logg, \fsep\ and \ps. Therefore, we expect higher contamination rates at lower Galactic latitude. We also expect our inference of the \ps\ to be highly sensitive to the error of the input SED given that the signal of an RC star is on the order of the typical uncertainty (see Figure \ref{fig:synth}). Evaluating the contamination rates with our validation sample, in Figure \ref{fig:cont} we show that indeed the contamination rate steeply increases with larger G band uncertainties and lower Galactic latitude.  

Given that our data has a slightly different distribution of Galactic latitudes and G band errors than our validation sample, the contamination rate and completeness for a given sample may be different than that calculated by the validation sample. To account for this, we calculate our final sample contamination rate by integrating the contamination rate as a function of Galactic latitude and fractional G band uncertainty multiplied by the normalized number density of our final sample. This leads to a contamination rate of $\sim$ 20\% for the ``tier 1" sample and $\sim$ 35\% for the ``tier 2" sample. This is an increase  of $\sim$ 3-5\% compared to the validation sample. However, as shown in Figure \ref{fig:cont}, a cut in either the fractional G band magnitude errors or Galactic latitude can lead to much lower contamination rates. For example, applying a cut of $|b|$ >30 degrees  to the ``tier 2" sample leads to a sample of 15,000 stars with an integrated contamination rate of $\sim$ 9\%.

\subsection{Galactic Distribution of Red Clump Sample}

 \begin{figure}
    \centering
    \includegraphics[width=\columnwidth]{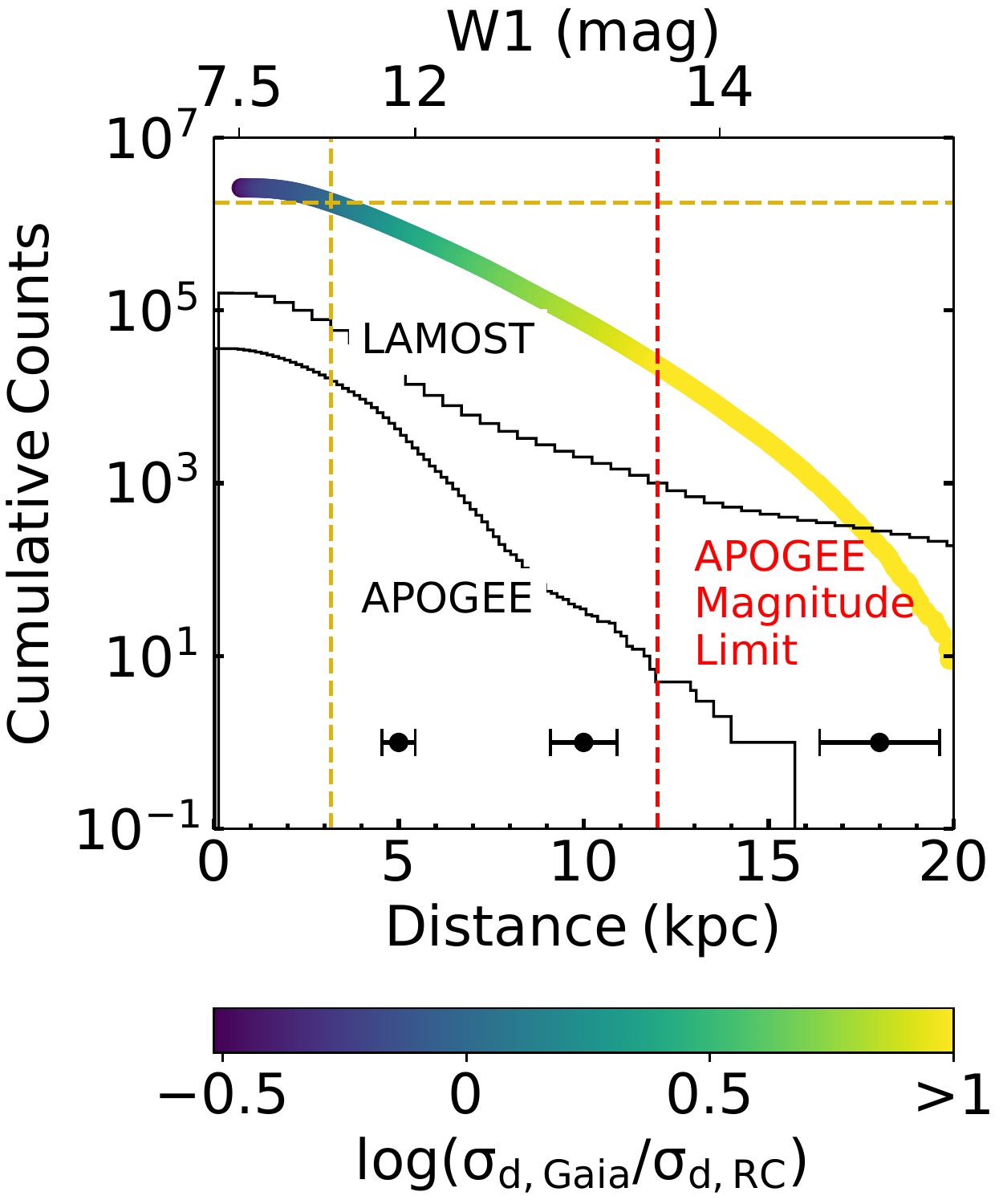}
    \caption{Cumulative distribution of distances for our sample of 2.6 million RC stars. We show the corresponding RC W1 magnitude on the top x-axis for reference. Our sample is colored by the ratio of $Gaia$ DR2 distance uncertainty to the distance uncertainties of RC stars. We also show the spectroscopic LAMOST and APOGEE RC samples from \citet{Ting2018} in black for comparison. The red dashed line corresponds to the RC distance for the APOGEE magnitude limit (H=14 mag). The yellow dashed lines show where the $Gaia$ DR2 uncertainties become larger than the RC distance uncertainties.  Our RC distances are more precise at distance $>$ 3 kpc. The $Gaia$ uncertainties degrade faster because they scale as distance squared rather while the RC distance uncertainties are linear with distance. The black error bars show the typical RC distance uncertainty for distances of 5, 10 and 18 kpc. }
    \label{fig:dists}
\end{figure}
 
 \begin{figure*}
    \centering
    
    \includegraphics[width=\linewidth]{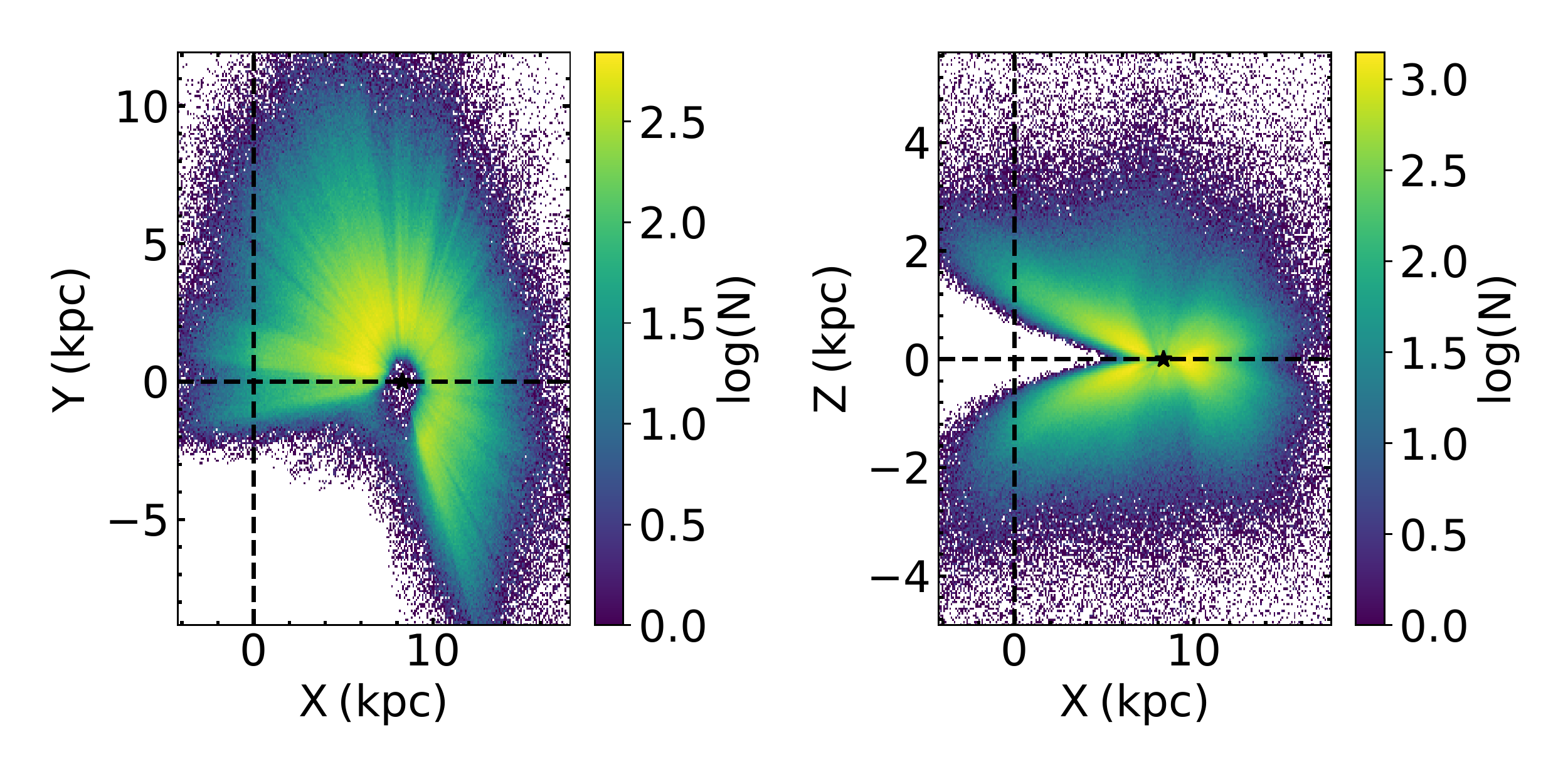}
    \caption{Galactic distribution of our sample of 2.6 million RC stars. On the left is the density of stars in Galactic coordinates, X and Y with |Z| < 3 kpc and on the right are the Galactic coordinates X and Z. The Galactic center is located at (0,0,0) in both plots, and the Sun is located at (8.3,0,0) kpc. The missing section of the disk is due to the Pan-STARRS footprint; Pan-STARRS does not reach declinations $<-30^{\circ}$. High levels of extinction prevent current photometric surveys from observing RC stars in bands bluer than the K band at the Galactic center. This observational effect could explain why we do not see the Galactic bar in the X-Y plane.}
    \label{fig:map}
\end{figure*}

Figure \ref{fig:dists} shows the reverse cumulative distribution of the derived heliocentric distances for the ``complete" sample (``tier 2"). For comparison, we show the LAMOST and APOGEE spectroscopic RC samples from \citet{Ting2018} in black. Even though the LAMOST sample is the largest and most distant sample of spectroscopically derived RC stars, our photometrically derived sample is significantly larger even at large distances. Therefore, our sample provides more precise distances where they matter, e.g., in the distant Galaxy where there are many open questions about its structure and where the distances are most difficult to derive. For comparison, we also derive uncertainties for distances derived using $Gaia$ DR2 parallaxes. For this we use the equation$^1$ $\sigma_{d}=\sigma_{\varpi}d^2$, where $\sigma_{\varpi}$ is the individual reported \gaia\ DR2 parallax uncertainty. In Figure \ref{fig:dists}, we color our distribution of stars by the ratio of the \gaia\ distance uncertainties to the typical uncertainty for RC stars ($\sim$ 9\%) at the given distance. The dark blue dashed lines indicate where distances measured using RC stars become more precise than distances provided by \gaia\ DR2 parallaxes. \textit{In total, our sample contains over 1.8 million stars with distances more precise than the distance given by $Gaia$ DR2 parallaxes (see Figure \ref{fig:dists})}.

We show the Galactic distribution of the sample of $\sim$ 2.6 million RC stars in Figure \ref{fig:map}.   Our sample reaches farther into the Galactic halo and Galactic center than previous pristine RC samples \citep{Bovy2014,Ting2018,Wu2019} with our most distant RC star at a heliocentric distance of $\sim$ 20 kpc. This maximum distance is consistent with the W1 limiting magnitude (W1 = 17.1 mag) and the absolute magnitude of RC stars in W1 ($\rm{M_{W1}}$ = -1.68 mag) which corresponds to a distance of 30.06 kpc. As shown in Figure \ref{fig:map}, extinction prevents us from observing the Galactic center. Assuming an $\rm{A_V}$ =30 mag at the Galactic center, in order to observe a RC star at that distance (8.3 kpc) we would need to use the infrared. A RC star at that distance and level of extinction is just on the edge of the magnitude limit of the 2MASS survey for the K band (K=14.3 mag). The only bands in which RC stars at the Galactic center are observable in current large surveys are the K, W1 and W1 bands. However, given the typical uncertainties and CN information content, these bands would not be sufficient to effectively select RC stars (see Figure \ref{fig:synth}). Future infrared photometric surveys (e.g., WFIRST) may provide sufficient depth and precision to select an RC sample in the Galactic center.

\section{Interpreting the Networks}
\label{sec:interpret}

\begin{figure*}
    \centering
    \includegraphics[width=\linewidth]{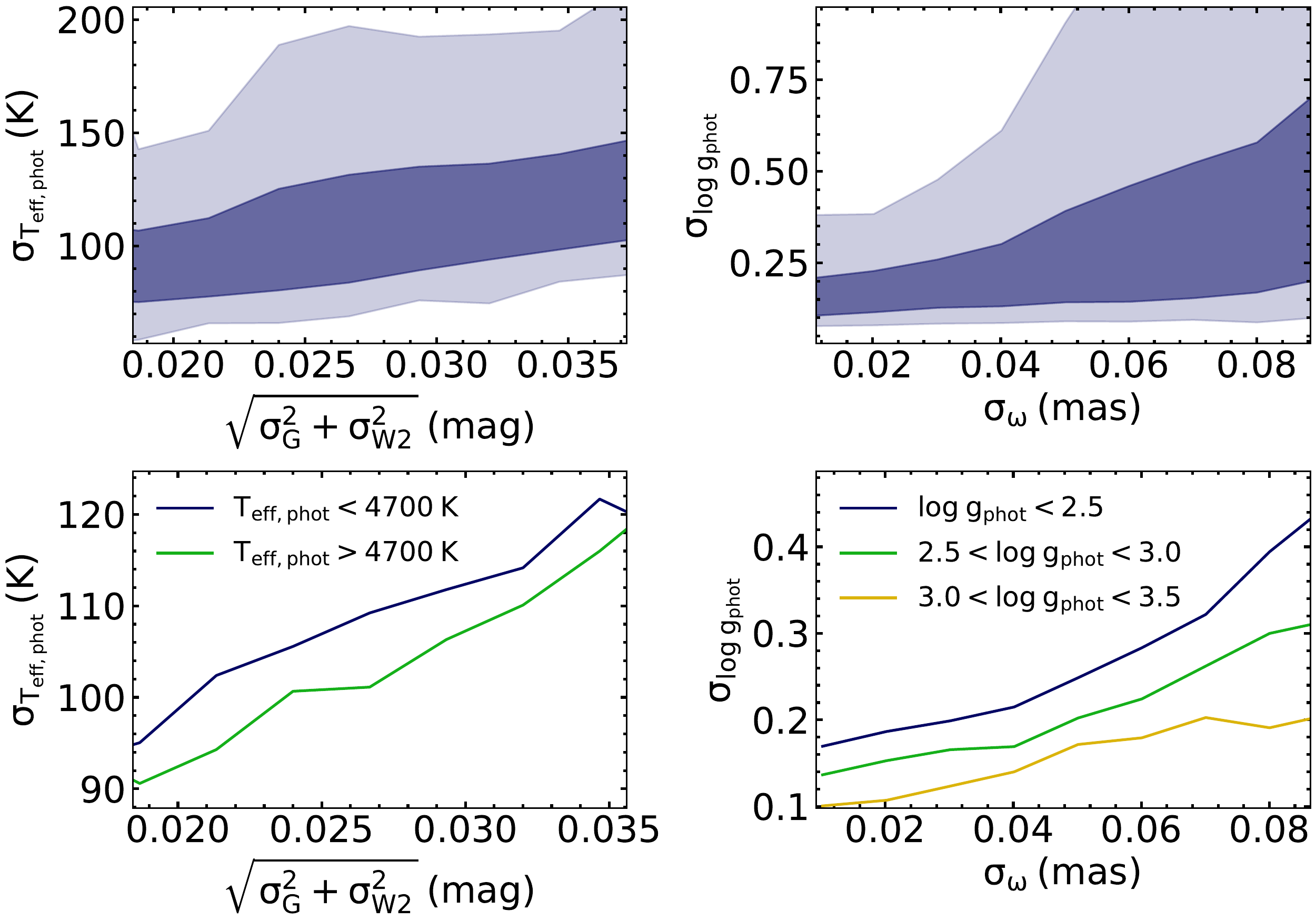}
    \caption{The uncertainties of inferred parameters \teff\ and \logg\ as a function of the uncertainties in the input data. Specifically we compare them to uncertainties in the specific inputs that are deemed crucial for the inference. For \teff\ this is the color of the star which can be described as G-W2.  For \logg\ it is the parallax ($\varpi$; see Section \ref{sec:int_t_g}).  The top plots show the 1$\sigma$ (dark blue) and 2$\sigma$ (light blue) intervals while the bottom plots show the median trend in different parameter ranges. The \logg\ uncertainties are larger for stars with \logg\ <2.5 dex likely because they are typically more luminous than higher \logg\ giants and therefore have larger fractional parallax uncertainties for a given parallax uncertainty. Similarly, stars with \teff\ <4700 K likely have higher \teff\ uncertainties because they are at the tip of the RGB and are brighter. Therefore, they are typically farther, leading to higher fractional uncertainties to the magnitude and the parallax.  }
    \label{fig:errs}
\end{figure*}

When using neural networks, it is often useful to understand what information is harnessed by the neural network to make the inference. This can be especially important when there might be confounding variables that the network could be using to make the inference instead of the desired physical mechanism. After all, neural networks aim to understand the correlations between variables. Domain knowledge is required to determine if the established correlation agrees with our understanding of the causal structure. In this section, we describe our attempts of trying to open the ``black box" and determine what correlations the network uses to infer the parameters.

\subsection{Interpreting the Reported Uncertainties}\label{sec:int_t_g}
One of the advantages to using a Mixture Density Network is that the inferred parameters also have uncertainties associated with them. The uncertainties are determined from the widths ($\sigma(x)$) of the output GMMs (see Section \ref{sec:MDN}). Although the network is not exposed to the input uncertainties, we find that the uncertainties on the inferred parameters correlate with the input uncertainties.

In the top panels of Figure \ref{fig:errs}, we show how the uncertainty on the inferred \teff\ and \logg\ relate to the uncertainties on the input parameters, which we suspect to be essential for the respective inference. The correlation shown in Figure \ref{fig:errs} suggests that the network is able to recognize a noisy SED and assign it a larger width of the GMM. Given that noise acts as a non-physical perturbation to the input, it is possible that the network detects these non-physical perturbations and during the training, learns it is unable to infer the parameter as accurately for SEDs with large perturbations as those with only small perturbations.

In the lower panels of Figure \ref{fig:errs}, we show the break down of these trends into different parameter bins. The lower left panels shows that giant stars with \teff\ $<$ 4700 K have higher \teff\ uncertainties for a given G and W2 uncertainty. This is likely because cooler stars are more luminous and due to the brightness limit of the surveys, luminous stars are often not observed when they are close to the Sun. As a consequence, cooler stars are statistically further way from their hotter counterparts. This leads to both a larger fractional uncertainty in photometry as well as parallax, which could explain why cooler stars have slightly larger uncertainties in their temperature estimates. Similarly, giants that are more evolved (lower \logg) are preferentially missed when they are very close by; as a consequence, they are biased toward larger distances. Therefore, they typically have a larger fractional uncertainty in parallax which causes are larger uncertainty in \logg.

\subsection{Parameter Importance for the \ps\ Inference} 
\label{sec:int_p}

We started this work with the conjecture that the network will pick up the difference in the [C/N] ratios between RC stars and red giant branch stars from the photometry and would use that information to determine the inferred \ps\ value. Indeed, the results shown in Figure \ref{fig:roc} demonstrate that we are more effective at picking RC stars using the \ps\ compared to just from the photometrically derived \teff\ and \logg. This indicates that the network is using more information than \teff\ and \logg\ to infer the \ps\ and therefore select RC stars.

\begin{figure*}
    \centering
    \includegraphics[width=\linewidth]{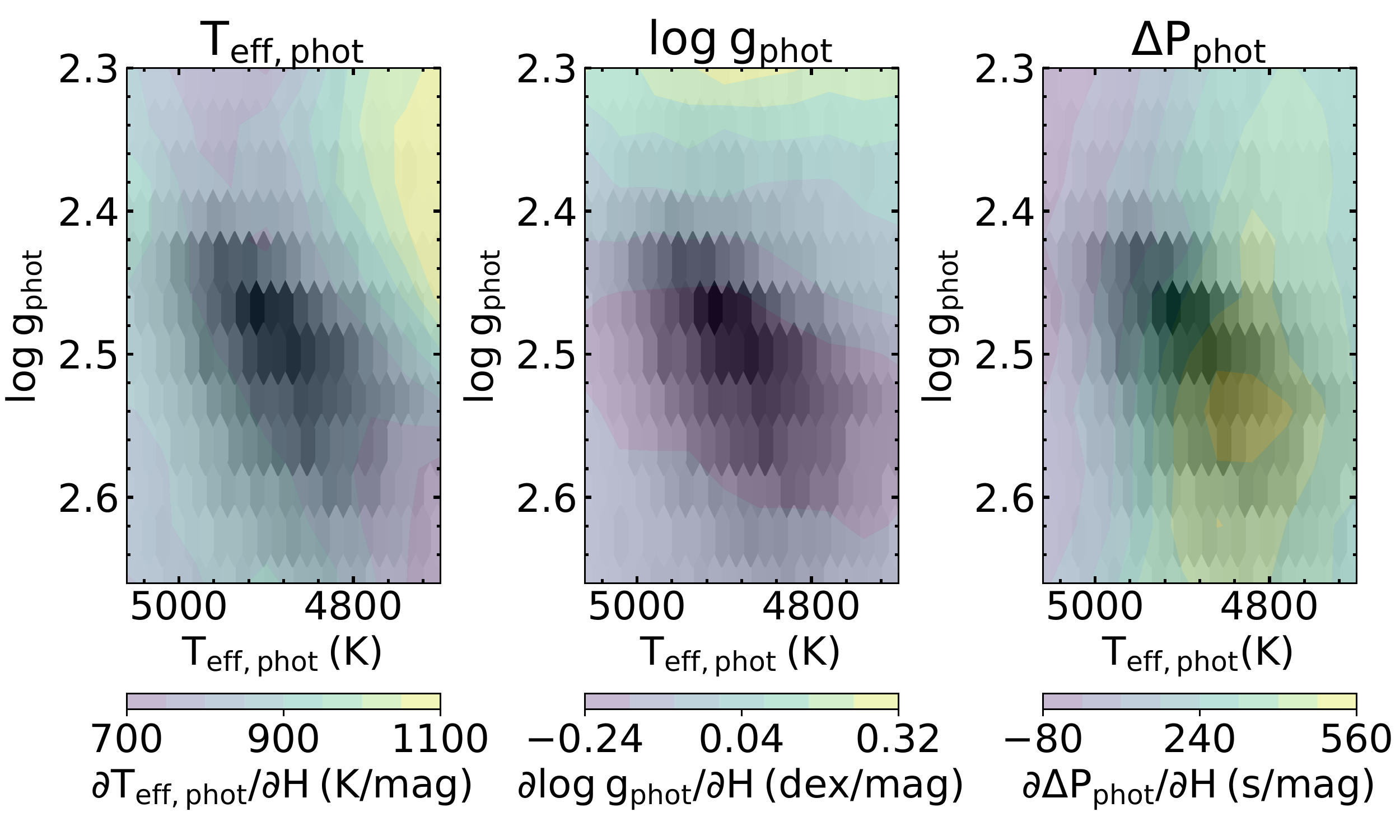}
    \caption{The empirical derivatives of \teff, \logg, and \ps\ with respect to the H-band as a function of \teff\ and \logg. In the background we show the counts of stars in grey hexagonal bins with the largest counts being the darkest. This is to show where the RC is in \teff\ and \logg\ and how it correlates with the derivatives. In the left and middle plot we see the \teff\ and \logg\ derivatives are smallest in the RC region while in the right plot the \ps\ derivatives are the largest in the RC region. The gradient of the derivative of \ps\ as a function of \teff\ and \logg\ is consistent with what is expected for a change in [C/N] due to mixing (See Figure \ref{fig:contours}), demonstrating that the network is learning the [C/N] ratio of RC stars.  }
    \label{fig:Hders}
\end{figure*}

In order to better understand the extra information the network is using to infer \ps, we calculate the empirical derivative of the inferred \ps\ with respect to each of the photometric bands. This is done by adding and subtracting 0.001 mag to the selected band for each SED in our testing set and re-inferring the \ps. Therefore, we have an empirical derivative calculated for giant stars with a range of \teff\ and \logg\ values, including both RC and RGB stars. We repeat this process with the \teff\ and \logg\ networks so that we have derivatives of \ps, \teff, and \logg\ with respect to each band. In Figure \ref{fig:Hders}, we show how the median derivatives of \teff, \logg\ and \ps\ with respect to the H-band vary with \teff\ and \logg. This is done by calculating the median of the validation data set for 100 (10$\times$10) \teff\ and \logg\ bins. We choose to show the H-band because we know it has CN molecular lines \citep{Hawkins2018} and therefore would be sensitive to any change in the [C/N] ratio (see Figure \ref{fig:synth}). 

We expect the network to have learned that if a star is outside the typical RC \teff\ and \logg\ parameter range, then it is likely not a RC star and it assigns that star a small and relatively constant \ps\ (\ps\ $\approx$ 80s), i.e., the derivative of \ps\ would be small. However, if the star is in the RC \teff\ and \logg\ range, then the network must determine if the SED is consistent with the expected change in [C/N] in order to infer the value of the \ps\ to assign. We expect that if the \ps\ network is using this information it should cause the derivative in the RC region to be large in order to be sensitive to the change in [C/N]. From Figure \ref{fig:Hders}, it is clear that the derivative of \ps\ with respect to the H-band has a local maximum near the RC region. Another important thing to note is that the variation of the derivative of \ps\ as a function of \teff\ and \logg\ is not similar to the variation showed for the derivatives of \teff\ and \logg. For example, both the derivatives \teff\ and \logg\ show local minima in the RC region. Additionally, the surfaces of constant derivative make up very different forms in the \teff\ and \logg\ space.  This gives further evidence that the \ps\ inference is based on more than simply the \teff\ and \logg.

 \begin{figure}
     \centering
     \includegraphics[width=\columnwidth]{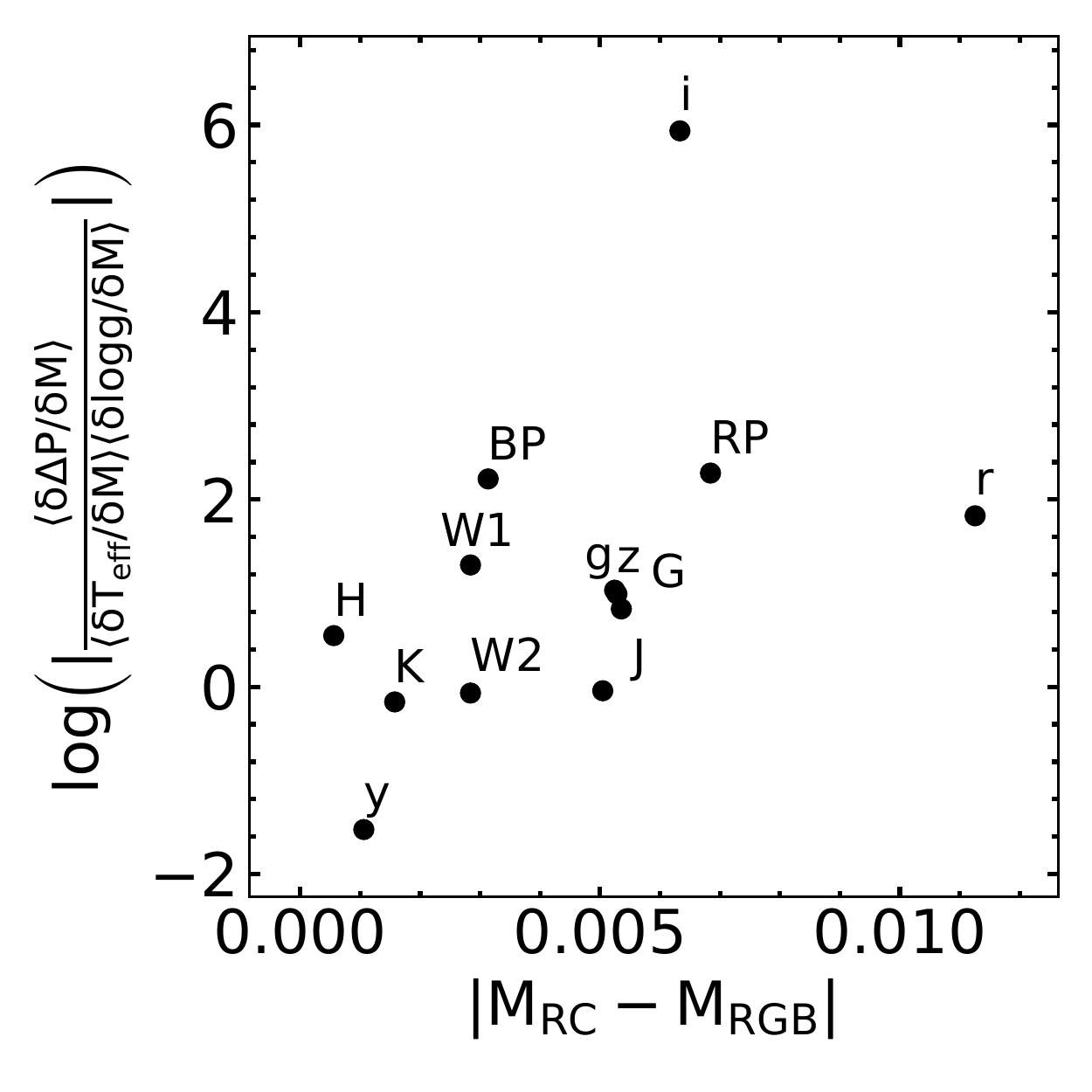}
     \caption{How the strength of the derivative of \ps\ with respect to each magnitude relates to the absolute expected change in the magnitude due to the CN mixing between an RGB and RC star. Specifically, on the y-axis, we plot the absolute value of the ratio of the derivative of \ps\ with respect to the magnitude to the derivative of \teff\ and \logg\ with respect to the magnitude. Here we assume the \teff\ and \logg\ are the same for both stars  (\teff\ = 4604 K and \logg\ = 2.62 dex)  but the RGB stars has a [C/N] =0 while the RC stars has [C/N] = -0.77 dex. The derivatives shown correspond to the median empirical derivative found for stars in the RC region (4100 K < \teff\ < 5100 K and 2.3 dex< \logg\ <2.7 dex). The bands with derivatives of \ps\ that are relatively larger than the derivatives of \teff\ and \logg\ generally also have CN molecular lines within them and therefore are more impacted by change in [C/N]. This indicates that the inference of \ps, and hence our RC selection, is likely aided by the imprint of a change in [C/N] on the SED. }
     \label{fig:dmag_ders}
 \end{figure}
 
  \begin{figure}
     \centering
     \includegraphics[width=\columnwidth]{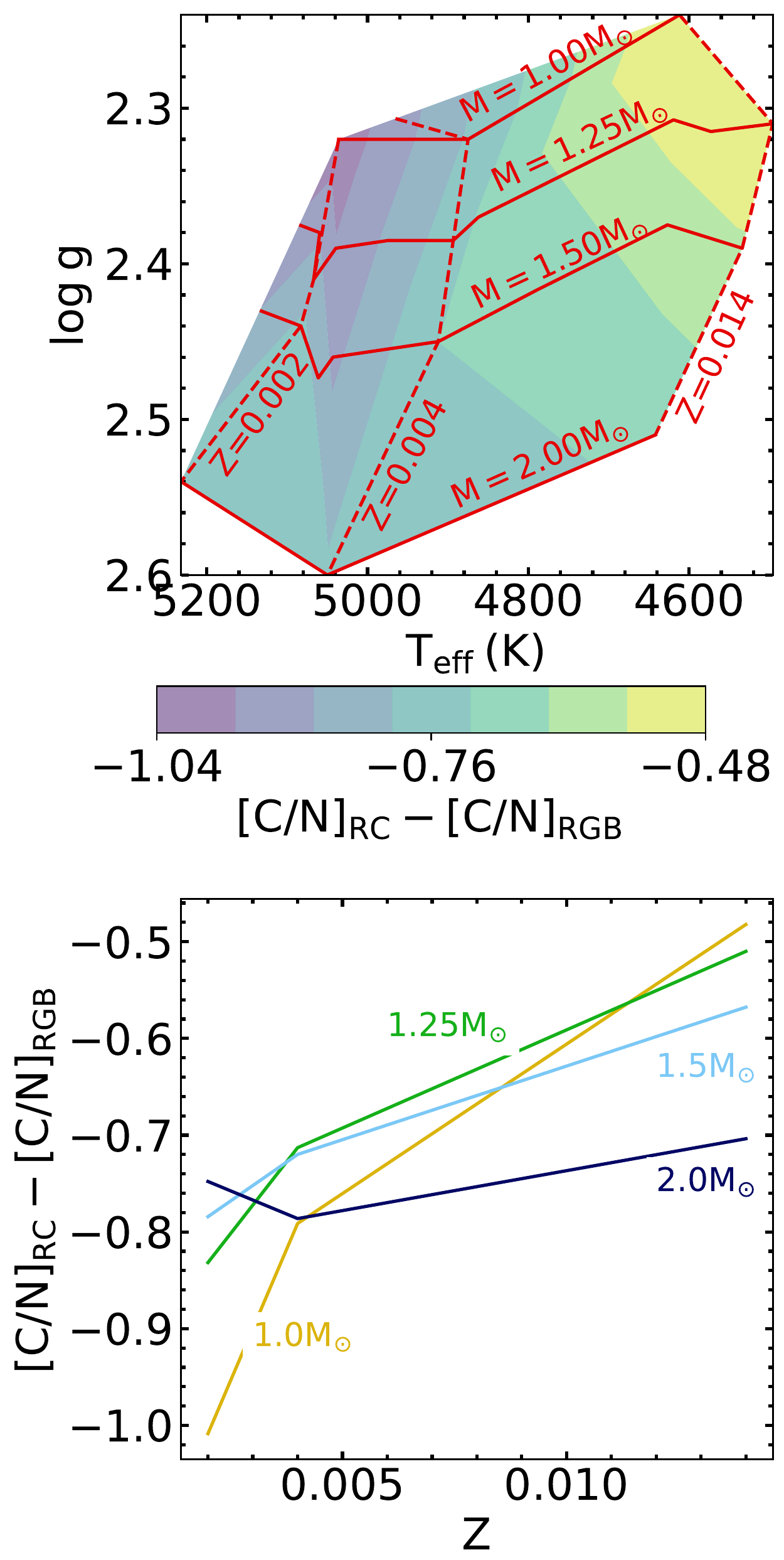}
     \caption{In the upper panel we show how the expected difference in [C/N] changes as a function of \teff\ and \logg. The models used are from \citet{Lagarde2012}. These differences are a result of stars with different initial masses and metallicities have different changes in [C/N] and different \teff\ and \logg\ during the core helium-burning phase. In the upper panel we show lines of constant mass in solid red lines and lines of constant metallicity in dashed red lines. As the mass of the star increases, so does the \logg. The \teff\ decreases with increasing metallicity. In general, the change in [C/N] becomes larger with decreasing metallicity. At the low mass end, this effect is very strong while at the high mass end there is almost no trend in [C/N] with metallicity. This is because the mass and metallicity both impact the depth of the convective zone during the RGB phase and therefore the amount of mixing that occurs.}
     \label{fig:cn}
 \end{figure}

  \begin{figure*}
     \centering
     \includegraphics[width=\linewidth]{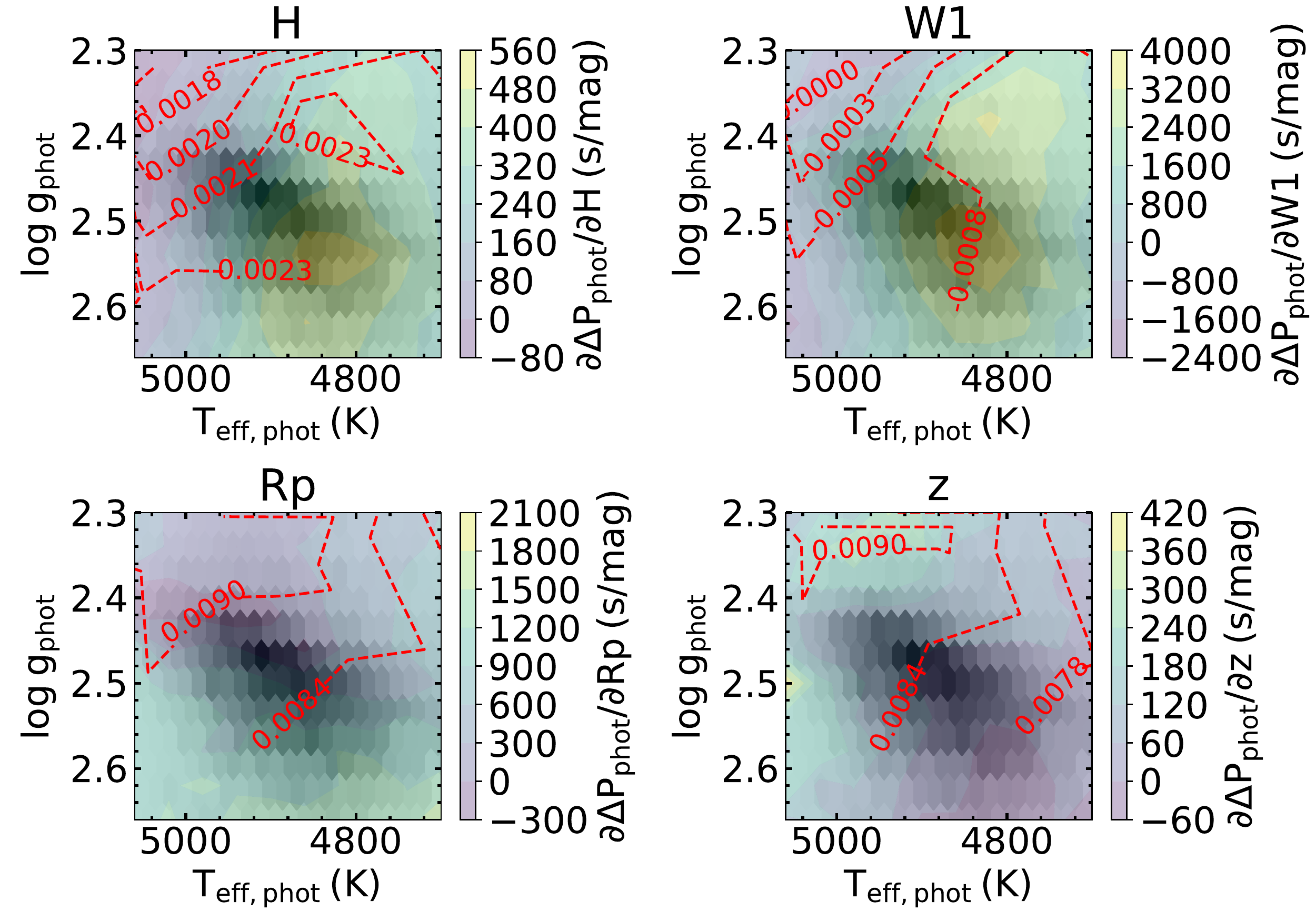}
     \caption{Showing the derivative of \ps\ with respect to the H, W1, Rp and z band as a function of \teff\ and \logg. The counts of stars are shown in hexagonal bins with darker bins corresponding to higher counts. The red dashed contours show the theoretical change in magnitude between a RC and RGB star due to the expected change in [C/N] for the selected band as a function of \teff\ and \logg. The theoretical change in magnitude contours are correlated with the empirical derivatives of \ps, indicating that the inference of \ps\ is based on the change in [C/N] between the RGB and RC. }
     \label{fig:contours}
 \end{figure*}

In addition to the derivatives, we calculate synthetic photometry to compare to the empirical results. If the network is using [C/N] information to infer \ps, we expect the derivative of the \ps\ to be correlated with the change in magnitude caused by a difference in [C/N] abundance.  In Figure \ref{fig:dmag_ders}, we show how the relative strength of the derivative of \ps\ relates to the calculated change in magnitude due to a change in [C/N] of -0.77 dex for each photometric band. The adopted change in [C/N] is motivated by models from \citet{Lagarde2012} and corresponds to an RC star with \teff\= 4604 K and \logg\ = 2.62 dex. The derivatives shown are calculated by taking the median of the derivatives of \ps, \teff, and \logg\ with respect to each band for stars with 4100 K < \teff\ < 5100 K and 2.3 dex< \logg\ <2.7 dex, the \teff\ and \logg\ of RC stars. We define the relative strength of the derivative of \ps\  as the ratio of the derivative of \ps\ with respect to the magnitude to the derivative of \teff\ and \logg\ with respect to the magnitude. We know that the \ps\ inference is modulated by the \teff\ and \logg\ of the star because the inference of \teff, \logg\ and \ps\ are weakly correlated. In addition, the empirical gradients for \teff, \logg, and \ps\ are each derived from separate networks, as such the correlation can manifest itself in the gradients. However, for these purposes, we are only interested in the sensitivity of the \ps\ inference to each band irrespective of a change in \teff\ and \logg. We take the absolute values of the derivative of \ps\ divided by the derivative of \teff\ and \logg\ in order to show which bands are most important for the \ps\ inference beyond their impact on the inferred \teff\ and \logg. From Figure \ref{fig:dmag_ders}, it is clear that the relative strength of the derivative of the \ps\ is correlated with the [C/N] information content in the selected photometric band. This indicates that the \ps\ inference is at least partially based on the change in [C/N] in the parameter range. 

We also calculate the expected change in magnitude due to a change in [C/N] as a function of \teff\ and \logg\ using synthetic photoemtry. To do this we first calculate the theoretical change in [C/H], and [N/H] as a star transitions from RGB to RC using models from \citet{Lagarde2012}. We calculate these changes for stars with a variety of initial masses (0.85, 1.0, 1.5 or 2.0 $\rm{M_{\odot}}$) and metallicities (Z= 0.002, 0.004 or 0.014) which correspond to a variety of \teff\ and \logg\ during the RC phase. Therefore, we have the change in [C/N] as a function of \teff\ and \logg\ for RC stars. We use these values to synthesize a grid of stellar spectra (see Section \ref{sec:data}), with RGB and RC [C/N] values. We then calculate the corresponding magnitudes for each photometric band (G, BP, RP, g, r, i, z, y, J, H, K, W1 and W2) using {\small PYPHOT}\footnote{https://mfouesneau.github.io/docs/pyphot} We calculate the expected difference in magnitude between in RGB and RC stars as a function of \teff\ and \logg\ for each photometric band. 

As shown in Figure \ref{fig:Hders}, the derivative of \ps\ with respect to the H-band is not constant over the RC region. This is expected given that the change in [C/N] is also not constant for all RC stars. The \teff\, \logg\ and [C/N] ratio of a RC star are a function of its initial mass and metallicity. The mass and metallicity of a star impact the depth of the convective zone during the RGB phase and therefore the amount of CN mixing. As shown in the lower panel of Figure \ref{fig:cn}, lower metallicity stars have a larger change in [C/N]. This effect is especially large for lower mass stars while stars with mass of 2 $\rm{M_{\odot}}$ have almost no trend in [C/N] with metallicity. The cumulative effect of this is shown in the upper panel of Figure \ref{fig:cn}, where the change in [C/N] is shown as a function of \teff\ and \logg. To expand on this further, we convert the change in [C/N] into a change in magnitude for selected bands using synthetic spectra. In Figure \ref{fig:contours},  we show how the theoretical changes in magnitude relate to the empirical derivatives of \ps\ with respect to the selected band as a function of \teff\ and \logg. The shape of the contours are highly correlated. This provide further evidence that the \ps\ network is learning these small differences in magnitude between RGB and RC stars caused by the change in [C/N] and using this to infer the \ps.

To summarize, we have found that the inferred output uncertainties are correlated with the input uncertainties which suggests the network can identify noisy data. In addition, the effectiveness of the selection when using the inferred \ps\ compared to using just \teff\ and \logg\ indicates that the \ps\ inference is derived from more than a correlation with \teff\ and \logg. Furthermore, we calculated empirical derivatives of the \ps\ with respect to each photometric band and found that they correlate with the expected difference in magnitude due to a change in [C/N]. In total, our investigation of the trained network suggests that the \ps\ inference uses the [C/N] information imprinted on the SED and therefore can be used to effectively select RC stars.

\section{Summary} \label{sec:summary}

RC stars are standard candles proven to provide more accurate distance measurements than end-of-mission \gaia\ parallaxes at distances > 3 kpc \citep{Ting2018}. However, identifying large pristine samples of RC stars has historically been difficult. Red giant branch stars can have the same \teff\ and \logg\ making it easy to mistake them as RC stars, especially when the \teff\ and \logg\ estimates have larger uncertainties, e.g., from low-resolution spectra or SED. The asteroseismic parameters \ps\ and \fsep\ clearly separate helium core-burning RC stars from inert core red giant branch stars. These parameters have only been derived for $\sim$2,000 giant stars given the difficulty of the measurement and the amount of time required for light curve observations. Furthermore, the asteroseismic sample is restricted to the Solar neighbourhood due to the magnitude limit. Recently, \citet{Hawkins2018} and \citet{Ting2018} demonstrated that the \ps\ and \fsep\ can be derived from single-epoch stellar spectra. Specifically, RC stars can be selected from the difference in the carbon to nitrogen ratio due to mixing that occurs at the top of the red giant branch.

Given that the information to effectively select RC stars are in the spectra, we expect the same information to be in the SED. However, from synthetic photometry, we find that this signal is on the order of the uncertainty in the photometry. Combined with the systematics that limit SED modeling, we resort to using data-driven models and inference with probabilistic neural networks to sort out the RC signal in the photometry. Specifically, we choose to use a Mixture Density Network which provide a probabilistic output rather than a single best-estimate. This allows us to provide uncertainty estimates for our inferred parameters. 

We selected RC stars from the $\sim$ 200 million stars which have photometry from 2MASS, AllWISE, \gaia, and Pan-STARRS. We derive the \teff, \logg, \fsep, and \ps\ of these stars from 13 bands of photometry and parallax using a mixture density network. Comparing to external validation samples, we achieve accuracies to the level of 153 K, 0.32 dex, 80s, and 1.64 \michz\ for \teff, \logg, \fsep\ and \ps, respectively.  We achieve a contamination rate of $\sim$20\% for a sample of 405,000 RC stars (``tier 1"). This is similar to the contamination rate found when selecting RC stars using \teff\ and \logg\ derived from low-resolution spectra. We also present a catalog of 2.6 million RC stars with a contamination rate of $\sim$33\% (``tier 2"). We find that extinction and the input errors impact the contamination rate. We can obtain a contamination rate of $\sim$ 9\% for 15,000 stars with |$b$| >30 degrees. We derive distances with uncertainties of $\sim$9\% for our sample and show the Galactic distributions. Our sample reaches deep into the Galactic bulge and halo with the most distant stars at a distance of $\sim$ 20 kpc. Therefore, this sample has vast scientific potential for Galactic studies. 

We also attempt to interpret the neural networks. We find they are consistent with our expectations from stellar evolution models and synthetic spectral models. Our study is an example showing that combining domain knowledge with the power of inference of machine learning can be a useful combination. This can be used to avoid machine learning algorithms from blindly exploiting unphysical correlations in the data which can cause unknowingly biased results. By calculating empirical derivatives of the \ps\ and comparing them with model expectations in the change in [C/N], including the effects of mass and metallicity,  we conclude that our network harnesses the difference in the [C/N] ratio to determine the \ps, along with the \teff\ and \logg. 

Increased precision photometric data will help to reduce the contamination. Specifically, WFIRST, Euclid and the Rubin Observatory should all provide improved photometry, especially in the infrared where 2MASS and AllWISE are currently on the order or 0.025 mag uncertainties which is larger than the expected magnitude change in these bands. In addition, these surveys will provide deeper magnitude limits and therefore will allow us to expand our sample to more distant RC stars. This will especially be useful in the Galactic center where extinction currently prevents us from seeing. Last, a larger and less contaminated training set will also improve the results. It would be ideal if we could train out network on asteroseismically derived parameters rather than spectroscopically derived. However, the asteroseismic sample is currently too small. Larger asteroseismic catalogs with derived \ps\ and \fsep\ values, e.g. with Plato and TESS, are needed to improve the training.

\section*{Acknowledgements}
{\small
This project was initiated in the Kavli Summer Program in Astrophysics held at UC Santa Cruz in 2019, funded by the Kavli Foundation and by the National Science Foundation. We thank them for their generous support.

ML acknowledges support from the Wootton Center for Astrophysical Plasma Properties under the United States Department of Energy collaborative agreement DE-NA0003843.

YST is grateful to be supported by the NASA Hubble Fellowship grant HST-HF2-51425.001 awarded by the Space Telescope Science Institute.

NSR is partially supported under the U.S. Department of Energy contract DE-AC02- 06CH11357 for the work at Argonne National Laboratory. 

KH has been partially supported by a TDA/Scialog grant funded by the Research Corporation and a Scialog grant funded by the Heising-Simons Foundation. KH acknowledges support from the National Science Foundation grant AST-1907417.

The authors acknowledge the Texas Advanced Computing Center (TACC) at The University of Texas at Austin for providing high performance computing resources that have contributed to the research results reported within this paper. URL: http://www.tacc.utexas.edu

This work has made use of data from the European Space Agency (ESA)
mission {\it Gaia} (\url{https://www.cosmos.esa.int/gaia}), processed by
the {\it Gaia} Data Processing and Analysis Consortium (DPAC,
\url{https://www.cosmos.esa.int/web/gaia/dpac/consortium}). Funding
for the DPAC has been provided by national institutions, in particular
the institutions participating in the {\it Gaia} Multilateral Agreement.

This publication makes use of data products from the Two Micron All Sky Survey, which is a joint project of the University of Massachusetts and the Infrared Processing and Analysis Center/California Institute of Technology, funded by the National Aeronautics and Space Administration and the National Science Foundation.

The Pan-STARRS1 Surveys (PS1) have been made possible through contributions of the Institute for Astronomy, the University of Hawaii, the Pan-STARRS Project Office, the Max-Planck Society and its participating institutes, the Max Planck Institute for Astronomy, Heidelberg and the Max Planck Institute for Extraterrestrial Physics, Garching, The Johns Hopkins University, Durham University, the University of Edinburgh, Queen's University Belfast, the Harvard-Smithsonian Center for Astrophysics, the Las Cumbres Observatory Global Telescope Network Incorporated, the National Central University of Taiwan, the Space Telescope Science Institute, the National Aeronautics and Space Administration under Grant No. NNX08AR22G issued through the Planetary Science Division of the NASA Science Mission Directorate, the National Science Foundation under Grant No. AST-1238877, the University of Maryland, and Eotvos Lorand University (ELTE).
}
\bibliography{bibliography}
\bsp	
\label{lastpage}
\end{document}